\documentclass[acmsmall]{acmart}

\AtBeginDocument{%
  }

\setcopyright{acmlicensed}
\copyrightyear{2025}
\acmYear{2025}
\acmDOI{XXXXXXX.XXXXXXX}

\usepackage{graphicx}
\usepackage{multirow}
\usepackage{colortbl}
\usepackage{array}
\usepackage{subcaption}
\usepackage[flushleft]{threeparttable}
\usepackage{float}
\usepackage{longtable}
\usepackage{caption}
\usepackage[most]{tcolorbox}
\DeclareCaptionFont{blah}{\small\sffamily}
\def\tsc#1{\csdef{#1}{\textsc{\lowercase{#1}}\xspace}}
\tsc{WGM}
\tsc{QE}
\tsc{EP}
\tsc{PMS}
\tsc{BEC}
\tsc{DE}

\acmJournal{JACM}
\acmVolume{37}
\acmNumber{4}
\acmArticle{111}
\acmMonth{8}

\begin{document}

\title{Trapped by Expectations: Functional Fixedness in LLM-Enabled Chat Search}


\author{Jiqun Liu}
\authornote{Corresponding author.}
\email{jiqunliu@ou.edu}
\orcid{0000-0003-3643-2182}
\affiliation{%
  \institution{The University of Oklahoma}
  \streetaddress{401 W Brooks Street}
  \city{Norman}
  \state{OK}
  \country{USA}
  \postcode{73019}
}

\author{Jamshed Karimnazarov}
\email{jamshed.k@ou.edu}
\affiliation{%
  \institution{The University of Oklahoma}
  \streetaddress{110 West Boyd Street}
  \city{Norman}
  \state{OK}
  \country{USA}
  \postcode{73019}
}

\author{Ryen W. White}
\email{ryenw@microsoft.com}
\affiliation{%
  \institution{Microsoft Research}
  \streetaddress{14820 NE 36th Street}
  \city{Redmond}
  \state{WA}
  \country{USA}
  \postcode{98052}
}



\begin{abstract}
\textit{Functional fixedness}, a cognitive bias that restricts users' interactions with a new system or tool to expected or familiar ways, hinders the full potential of \textit{Large Language Model (LLM)-enabled chat search}, especially in complex and exploratory tasks. To investigate the impact of functional fixedness, we conducted a crowdsourcing study involving 450 participants, each completing one of six decision-making tasks spanning topics such as public safety, diet and health management, sustainability, and AI ethics. Participants were asked to engage in a \textit{multi-prompt} conversation with ChatGPT for addressing the task, allowing us to compare pre-chat \textit{intent-based expectations} with observed interactions. We found that: 1) Several aspects of pre-chat expectations on chat search are closely associated with users' prior experiences with ChatGPT, search engines, and virtual assistants; 2) Users' prior system experience shapes their language use and prompt reformulation strategies, with frequent ChatGPT users reducing deictic terms and hedge words and frequently adjusting prompting strategies, users with rich search experiences maintaining structured and less-conversational queries and minimally modifying the prompts, and users who used virtual assistants frequently favoring directive, command-like prompts, reinforcing functional fixedness in chat adaptability; 3) When the system failed to meet expectations, participants adjusted their strategies by generating more detailed prompts with increased linguistic diversity, reflecting adaptive tactic shifts. These findings indicate that while preconceived expectations initially constrain users, unmet expectations create motivations for behavioral adaptation and, with appropriate system interventions, may encourage broader exploration of LLM functionalities. In addition, this research establishes a comprehensive typology for characterizing \textit{user intents} in chat search. Mitigating functional fixedness may not only enhance task performances but also extend the use of LLMs beyond routine queries to more creative and analytical scenarios. Our research provides a psychology-informed foundation for designing \textit{in-situ} user education to foster more engaging, productive, and inspiring LLM interactions.
\end{abstract}




\maketitle

\section{Introduction}
Large Language Model (LLM) applications, such as ChatGPT and Copilot, are reshaping the field of Information Retrieval (IR) by enabling conversational and generative interactions that move beyond traditional keyword-based document retrieval and directive virtual assistants~\cite{najork2023generative, white2025information, zhu2023large, kanoulas2025agent}. These systems offer users the opportunity to engage in multi-turn dialogues that can accommodate contextual and flexible queries and support complex, open-ended informational tasks. However, the potential of LLM-enabled chat interactions is often constrained by users’ preconceived mental models and associated expectations, domain knowledge and querying skills, as well as behavioral patterns. These constraints arise from cognitive biases, such as \textit{functional fixedness}, which limit users’ ability to explore the full range of capabilities provided by LLM systems~\cite{azzopardi2024search, liu2024decoy}. Functional fixedness, originally studied in the context of problem-solving psychology, refers to a tendency to focus on familiar uses or functionalities of a tool based on existing beliefs, knowledge, and expectations \textit{transferred} from similar tools or remembered experiences~\cite{adamson1952functional, german2005functional, munoz2018functional}. When users interact with LLMs, this bias restricts their ability and motivation to adapt their strategies, hinders creativity and exploration of new actions, and thus reduces the overall effectiveness of the interaction.

\begin{figure}[H]
    \centering
    \includegraphics[width=0.65\linewidth]{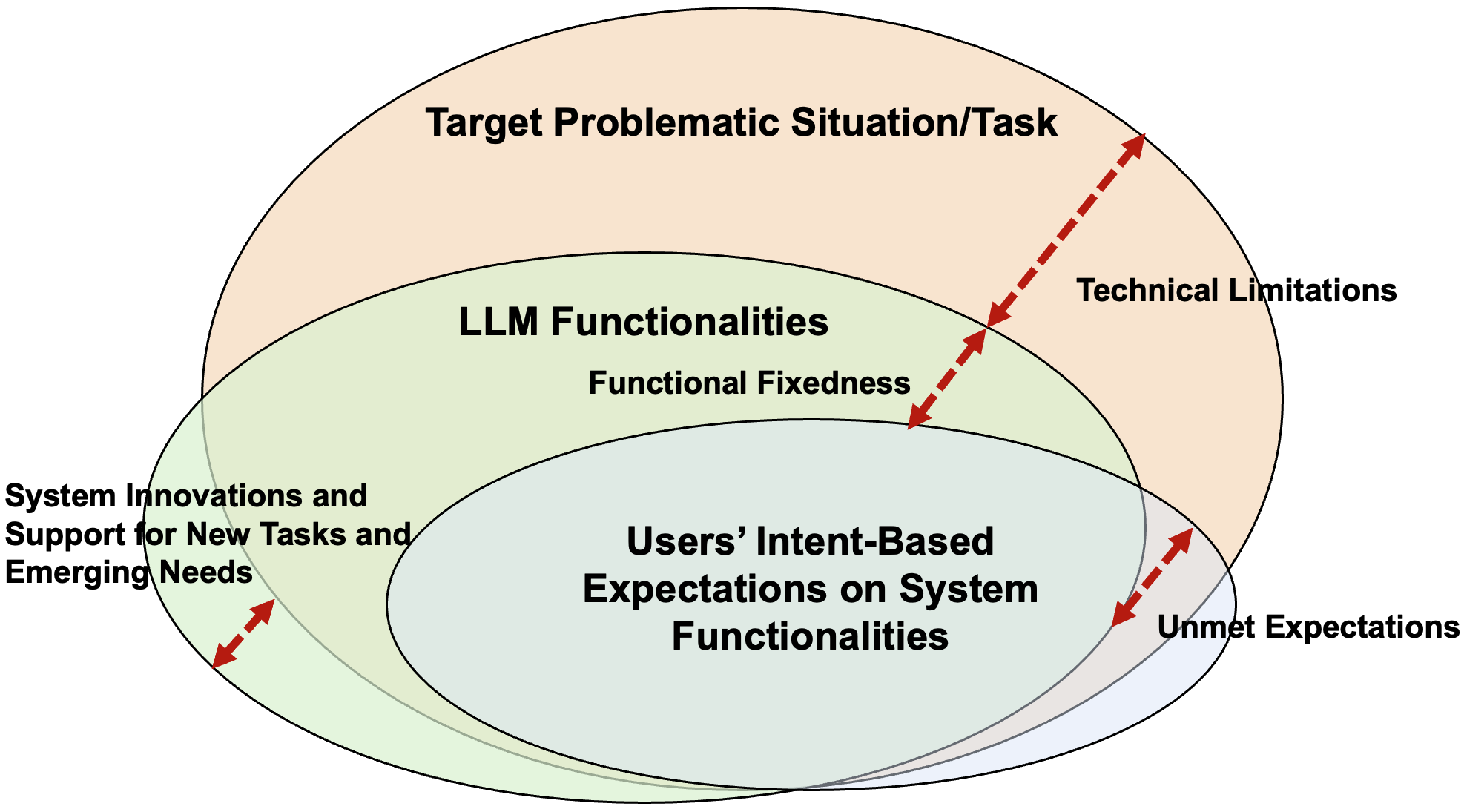}
    \caption{Functional Fixedness in LLM-enabled Chat Interactions. Note that this figure is only for illustrative purposes and is not to scale.}
    \label{fig:ff}
\end{figure}

The problem of functional fixedness is particularly significant in the context of LLM-enabled chat systems due to their expansive technical capabilities, unfamiliar interaction patterns that differ from query-driven search and Search Engine Result Page (SERP)-based browsing, and the implicit demands they place on users to adapt their mental models and explore unknown prompting strategies. Unlike traditional search engines, which are largely optimized for keyword-based text retrieval~\cite{buttcher2016information}, or virtual assistants, which handle narrowly scoped tasks or structured dialogues~\cite{zamani2023conversational, gao2020recent}, LLMs possess generative capabilities that allow them to reason and make judgments, synthesize information, and engage in complex and exploratory conversations. These systems are inherently designed to support creative problem-solving, flexible decision-making, and the articulation of multifaceted queries. However, the full realization of these capabilities depends on users’ ability to break free from fixed, preconceived notions about what these systems \textit{can} and \textit{should} do. When users remain anchored to interaction patterns derived from their experiences with simpler tools, they often do not take advantage of the advanced functionalities of LLMs, leading to suboptimal task outcomes and decreased satisfaction. Figure~\ref{fig:ff} illustrates this challenge in an LLM interaction context. While it is critical to address the technical challenges and enhance system's functionalities in problem-solving and task support, researchers and engineers should also examine the implicit impact of functional fixedness and enhance the alignment between users' mental models with systems' actual capabilities. 

The challenge of functional fixedness is not merely a matter of user familiarity but reflects deeper cognitive and behavioral tendencies explored in Human-Computer Interaction (HCI) and psychology communities~\cite{chrysikou2016functional}. Functional fixedness has long been studied in the context of problem solving, where it is shown to constrain individuals’ ability to perceive novel applications of tools or objects~\cite{munoz2018functional, camarda2018neural}. A large body of HCI research highlights how cognitive biases influence user interaction with technologies, shaping task performance, user interaction experience, and the adoption of new systems~\cite{santhanam2020studying, dingler2020workshop, liu2023behavioral}. Functional fixedness, in particular, becomes more pronounced when users engage with novel systems, as these contexts demand cognitive flexibility. Theories of cognitive load and mental model formation further emphasize how users rely on existing schemas and approaches to minimize effort when interacting with unfamiliar tools~\cite{skulmowski2022understanding, buchner2022impact}. These schemas, while efficient and helpful for addressing cold-start problems in some contexts, can act as barriers to exploring new functionalities or adapting to dynamic systems~\cite{zhangfalling}. Addressing these barriers is critical for unlocking the potential of LLMs, particularly since these tools and systems are increasingly being deployed in critical domains that require higher-order cognitive engagement, such as learning and education, healthcare, and professional decision-making~\cite{moore2023empowering, nazi2024large, cascella2023evaluating}.

On a technical level, LLMs themselves exacerbate the challenges posed by functional fixedness. While their generative capabilities allow for highly adaptive responses, these systems do not always provide transparent feedback about their full range of capabilities. Users may misinterpret ambiguous or incomplete system responses as inherent limitations, reinforcing their preconceived notions and expectations about what the system can achieve. Furthermore, the absence of explicit guidance or scaffolding in many LLM interfaces leaves users reliant on trial-and-error interactions, which could deepen frustration and discourage further exploration. This highlights the need for systems that actively support users in overcoming functional fixedness and other cognitive biases through automated mechanisms~\cite{liu2023behavioral}, such as affordance signaling, contextual nudges, and adaptive feedback. While researchers have examined cognitive biases on both the human and system sides~\cite[e.g.,][]{chen2024ai, chen2024decoy, liu2023behavioral, wang2024cognitively, lajewska2024can}, how functional fixedness affects users' interactions with LLM-based systems still remains understudied. This challenge hinders the effective use of LLM-enabled chat search systems and limits the alignment between users' mental models and LLM's evolving functionalities. 

To address the challenge above, our study investigates functional fixedness in LLM-enabled chat search by examining how users’ expectations, shaped by prior experiences, influence their interaction behaviors and task outcomes. We conducted a \textit{controlled crowdsourcing experiment} involving 450 participants, and adopted six diverse decision-making tasks across domains such as public safety, health and diet management, sustainability, and AI ethics. These tasks were designed to elicit multi-turn interactions and enabled us to study the adaptive strategies that users employed when system responses failed to meet their expectations. 

We found that functional fixedness constrains users’ ability to engage effectively with LLMs, particularly in tasks requiring creativity or analytical reasoning. However, the study also highlights the adaptive potential of users when faced with unmet expectations. Many participants responded to system failures by crafting linguistically diverse prompts, demonstrating that functional fixedness can be mitigated under certain conditions. These findings align with theoretical perspectives emphasizing the role of adaptive behaviors in overcoming cognitive biases~\cite{otuteye2015overcoming, rastogi2022deciding}. By designing systems that support such adaptation, it becomes possible to transform functional fixedness from a limiting factor into a catalyst for deeper exploration. Our research makes the following contributions:
\begin{itemize}
    \item This study systematically examines the influence of functional fixedness on user interactions with LLM-enabled chat search under varying tasks and \textit{in-situ} expectation states through a user experiment. The analysis identifies specific patterns in prompt formulation and adaptation that are shaped by users’ prior experiences with search, chatbots, and virtual assistants, and offers empirical evidence on how cognitive biases manifest in chat search.
    \item The study develops a comprehensive typology of user intents that characterizes diverse approaches to LLM interactions and adaptation processes. By bridging cognitive psychology and system design, the framework contributes to the theoretical understanding of user engagement in generative AI contexts and provides a foundation for user education.
    \item Methodologically, our work offers a robust study design for capturing, analyzing, and addressing functional fixedness and other related human cognitive biases in human-AI interactions at large scale. Future research in this area can adapt our experimental setup according to the specific research questions and the nature of biases being studied. 
    \item The findings suggest actionable strategies for mitigating the potential negative effects of functional fixedness, which can inform the design of next-generation LLM-enabled systems that promote effective user interactions while addressing cognitive constraints.
\end{itemize}

The structure of this article is as follows: Section 2 introduces previous research on functional fixedness, user intents and expectations, as well as information retrieval and conversational search; Section 3 presents our research questions (RQs); Section 4 provides an overview of crowdsourcing experimental setup; Section 5 describes experimental results of experiment; Section 6 summarizes our findings and responses to the proposed RQs and discusses some open challenges in this field.


\section{Background}
This section reviews and discusses existing theoretical background, empirical findings, and open challenges based on which we proposed our research questions and designed the study. 
\subsection{Functional Fixedness}
Functional fixedness is a cognitive bias that limits an individual's ability to use objects or concepts beyond their traditional functions, thereby impeding problem-solving, effective exploration, and creativity. Duncker and Lees~\cite{duncker1945problem} first illustrated this phenomenon through the "candle problem," where participants struggled to perceive a box of tacks as a potential candle holder, focusing instead on its conventional role as a container.~\citet{adamson1952functional} replicated and extended this work, and confirmed that prior exposure to an object's typical function hinders the ability to envision alternative uses. Subsequent research has explored functional fixedness across various problem-solving contexts. For instance, \citet{german2005functional} investigated its presence in non-industrialized cultures and found that even individuals with limited exposure to manufactured tools exhibit functional fixedness, which suggests its universality. \citet{chrysikou2016functional} examined the impact of pictorial examples on creative generation tasks, and showed that such examples can exacerbate functional fixedness by reinforcing standard uses of objects. In educational settings, functional fixedness poses challenges to learning.~\citet{furio2000functional} identified that students often struggle with chemistry problems due to rigid perceptions of molecular structures, a manifestation of fixation. \citet{mccaffrey2012innovation} proposed the ``\textit{generic parts technique} (GPT)'' as a method to overcome this bias and encourage individuals to decompose objects into separate constituent parts and consider their properties independently of their typical functions.

Neuroscientific studies have provided insights into the underlying mechanisms of functional fixedness. For example, \citet{camarda2018neural} utilized electroencephalography (EEG) to investigate neural correlates during creative idea generation, and found that overcoming functional fixedness is associated with increased activity in brain regions linked to cognitive flexibility. This aligns with findings by~\citet{chrysikou2011dissociable}, who reported that inhibiting the left prefrontal cortex can reduce functional fixedness. This result indicates that this brain region plays a role in maintaining conventional object representations. Interestingly, functional fixedness is not even limited to humans; it has been observed in non-human primates as well. \citet{voelter2017social} found that chimpanzees exhibit functional fixedness in tool use, often failing to repurpose tools for novel functions, indicating that this cognitive bias may have deep evolutionary roots. In the context of technology and design, functional fixedness can hinder innovation. \citet{youmans2014design} discussed strategies to mitigate this bias in engineering, such as promoting analogical thinking and encouraging the consideration of alternative functions during system design processes.

Recent studies have also explored the relationship between functional fixedness and other cognitive biases. \citet{herstatt2005use} examined how fixation effects can impede knowledge transfer in innovation processes, while~\citet{ollinger2008investigating} investigated the role of insight and restructuring in overcoming both functional fixedness and mental set biases during problem-solving. Understanding functional fixedness and developing strategies to overcome it are crucial, especially in AI application contexts, for better aligning user expectations with system functionalities and enhancing creativity across various domains, such as education, health, and innovation.

\subsection{Cognitive Bias in Information Retrieval}
Cognitive biases play a pivotal role in shaping user interactions with IR systems, and influences how information is sought, interpreted, and assessed~\cite{liu2023behavioral, azzopardi2024evaluating, liu2024decoy, liu2020investigating}. These biases manifest in ways that reinforce pre-existing beliefs, privilege certain sources over others, and ultimately affect decision-making, often without users being consciously aware of their influence~\cite{azzopardi2021cognitive, metzger2013credibility}. IR researchers found that users tend to engage with information selectively, forming queries that align with their expectations and interpreting results in ways that validate their prior knowledge~\cite{white2013beliefs, wang2024cognitively, suzuki2021characterizing}. The tendency to favor easily accessible or prominently ranked results, coupled with implicit trust in authoritative sources, contributes to information distortion and misinformation propagation~\cite{pan2021examination}.

This systematic filtering of information occurs at multiple levels within search processes, from initial query formulation to final selection of sources. Algorithmic rankings, designed to optimize relevance, often reinforce these biases by favoring content that is already widely accepted, thereby limiting exposure to alternative viewpoints~\cite{liu2023behavioral, lipani2019biases}. Users, in turn, exhibit cognitive tendencies that exacerbate these limitations, such as over-reliance on the first search results encountered, excessive trust in familiar sources, and susceptibility to the way information is framed~\cite{kahneman2003maps, metzger2013credibility, chen2022constructing}. 

Efforts to address biases in IR have focused on two key areas: system-level interventions and user-centered strategies. On the system side, \textit{algorithmic diversification} has been proposed as a means to expose users to a broader and fairer range of perspectives, counteracting tendencies toward selective exposure, biased clicking, and self-reinforcing search behaviors~\cite{santos2015search, maxwell2019impact, gao2022fair}. Transparency in ranking mechanisms, coupled with \textit{bias-aware} retrieval models, could further mitigate undue reliance on pre-established preferences~\cite{castillo2019fairness, chen2024decoy}. From a user perspective, behavioral interventions, such as adaptive ranking strategies and nudging techniques, have shown potential in encouraging more balanced engagement~\cite{liu2023toward, wang2024cognitively, wang2023investigating}. As IR systems continue to evolve, the challenge remains in balancing document relevance optimization with bias mitigation~\cite{liu2022matching, liu2022leveraging}. While biases are deeply ingrained in human cognition, integrating real-time feedback mechanisms and adaptive interfaces may provide new pathways for fostering critical engagement with retrieved results~\cite{liu2022matching}. Future research should refine bias-aware ranking techniques and further explore strategies that dynamically adjust search environments to promote more equitable interactions with retrieved or generated information~\cite{pan2021examination}. On the system side, researchers can also explore how LLM-enabled systems inherit and aggrevate biases from users in online information seeking~\cite{chen2024ai, echterhoff2024cognitive, liu2024decoy, he2025investigating}. 

\subsection{User Intent and Expectation in Information Retrieval}
User intent and expectation play a crucial role in shaping users' interactions with and evaluations on IR systems. While traditional IR studies classify intent into broad categories, such as informational, navigational, and transactional~\cite{broder2002taxonomy, rose2004understanding}, recent research has emphasized the complexity and evolution of user intent and cognitive states in interactive and multi-turn search scenarios~\cite{marchionini2006exploratory, liu2019task, zamani2023conversational, rha2016exploring}. User expectations, which are often built upon individual users' prior interaction experiences and in-situ intents, influence how they frame queries, evaluate retrieved content, and adjust their search strategies in sessions~\cite{wang2023investigating, wang2024understanding}. Prior experiences with search engines, virtual assistants, and generative AI systems shape users' pre-established and in-situ search expectations, and lead users to form mental models about how IR systems \textit{should} behave under queries of varying types~\cite{ingwersen2005information, liu2024search}. A mismatch between user expectations and search system performance can lead to frustration, while well-aligned interactions enhance engagement, satisfaction, and task performances~\cite{liu2023behavioral, wang2023investigating, wang2024task}.

In complex search tasks, user intent often shifts as users refine their queries and engage with retrieved information~\cite{liu2019task, liu2020identifying, markwald2023constructing}. Differing from static keyword-based queries, conversational and interactive search settings require iterative refinements, intent clarification mechanisms, and adaptive system responses~\cite{marchionini2006exploratory, kelly2015development, liu2020identifying}. Studies have shown that cognitive biases related to functional fixedness, such as anchoring and confirmation bias, can hinder users from effectively adapting their search behavior and affect their assessments on retrieved documents and overall search experiences~\cite{liu2023behavioral, azzopardi2021cognitive, white2013beliefs, ji2024towards}. When expectations are not met, users often react by modifying queries and prompts, seeking external validation, or disengaging entirely~\cite{wang2023investigating, edwards2017engaged, hassan2014struggling, feild2010predicting}. Systems that provide explicit feedback, transparency mechanisms, and dynamic recommendations could mitigate these challenges by breaking users' functional fixedness and guiding them toward more effective search and evaluation strategies~\cite{wang2024cognitively, liu2023toward}. Recent IR research has discussed and explored the possibility of integrating user intent and expectation modeling to improve search experiences and evaluations. Real-time intent recognition, interactive ranking adjustments, and adaptive interfaces have been proposed to enhance alignment between system outputs and user needs~\cite[e.g.,][]{zamani2023conversational, qu2018analyzing, zhang2018towards, wang2024task}. Future work should further explore how personalized interventions and \textit{expectation-aware} systems can optimize user performance in generative search and AI-mediated environments. 

\subsection{Summary}
The sections above highlight unresolved challenges in understanding how functional fixedness shapes user interactions in traditional IR activities and LLM-enabled chat search. While prior research has identified biases (e.g., confirmation and anchoring biases, decoy effect, reference dependence) as constraints on adaptive search behaviors, existing IR models provide limited mechanisms to mitigate these issues, especially in multi-turn conversations~\cite{chen2023reference, liu2023toward, azzopardi2021cognitive}. Additionally, user intents and expectations, shaped by past experiences with search engines and AI systems, play a critical role in query formulation, document judgment, and search tactic adaptation~\cite{liu2019task}, yet their interplay with LLM-driven retrieval remains largely underexplored. Addressing these gaps requires empirical research that systematically examines the impact of user expectations on prompting strategies, search refinement, and task performance. Our crowdsourcing experiment is conducted to provide new insights into these cognitive constraints and offer evidence-based recommendations for facilitating more flexible, creative, and intellectually engaging chat interactions.

\section{Research Questions}
To address the open challenges discussed above, our study investigates the impact of functional fixedness at both intent/expectation level and interaction level, and aims to answer following RQs:

\begin{itemize}
    \item \textbf{RQ1}: How are users' pre-chat intent-based expectations associated with their prior interaction experiences with chatbot, search engine, and virtual assistants?
    \item \textbf{RQ2}: During the interactions, how are users' chat behaviors associated with their prior interaction experiences with chatbot, search engine, and virtual assistants and task types?
    \item \textbf{RQ3}: To what extent are users' chat interactions and whole-conversation experiences associated with their in-situ expectation confirmation states?
    
\end{itemize}

\begin{figure}[t]
    \centering
    \includegraphics[width=1\linewidth]{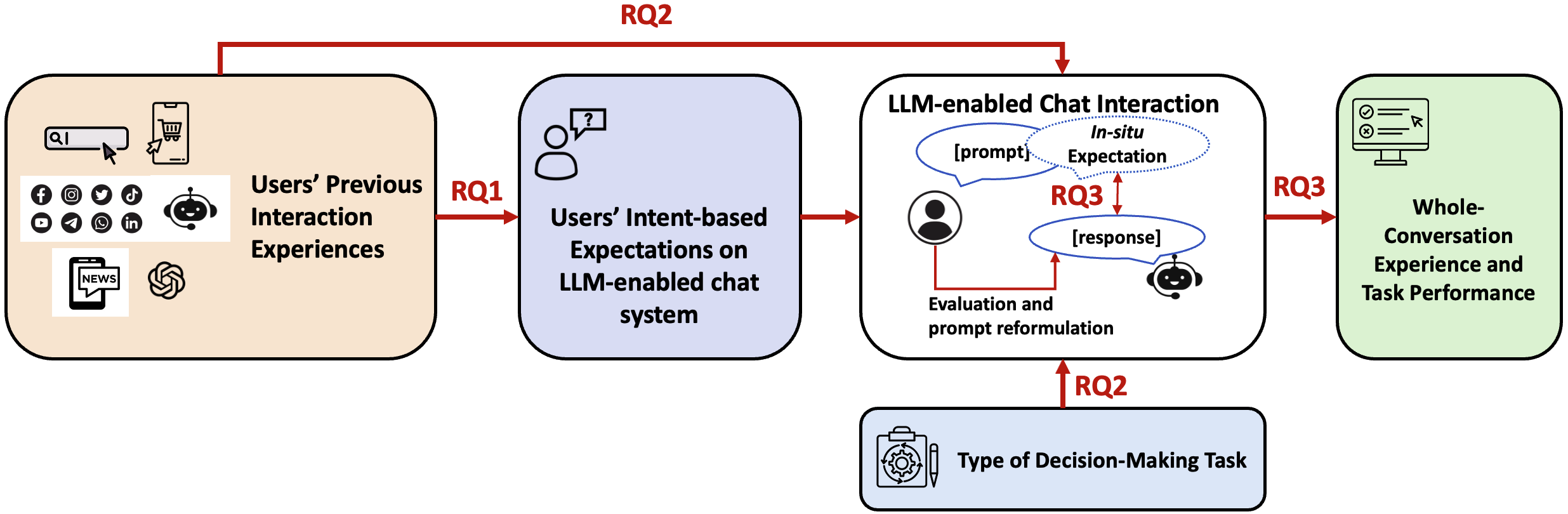}
    \caption{Functional Fixedness in LLM-enabled Chat Search: Research Questions}
    \label{fig:rq}
\end{figure}

RQ1 and RQ2 explore how users' expectations and behaviors are "fixated" by previous experiences with similar systems, which are the main focus on our research. RQ3 examines how users' expectation confirmation states (e.g., met, exceed, unmet) affect their functional fixedness status and adaptive prompting strategies. Figure \ref{fig:rq} presents the structure of our research problem. The following section introduces the experiment we designed for addressing the three RQs and for enhancing our understanding of the functional fixedness phenomenon in LLM-enabled chat search. 

\section{Methodology}
To answer the RQs, this study employs a controlled crowdsourcing experiment to examine the effects of functional fixedness. We investigate how users’ expectations, shaped by their prior experiences with search engines, virtual assistants, and chatbots, influence their prompting strategies and task performance. The study follows a multi-stage experimental design to systematically capture pre-chat expectations, in-situ interaction behaviors, and post-task experience assessments. This study adopted ChatGPT as an example LLM tool for interaction sessions due to its widespread adoption across diverse work and everyday life scenarios, making it a representative case for examining human-AI interaction patterns. As one of the most publicly accessible and extensively integrated generative AI systems, ChatGPT provides a practical testbed for studying how users engage with LLMs in real-world contexts, from information-seeking to creative ideation and decision-making. Its widespread use also allows for meaningful comparisons with emerging AI systems, which offers insights into evolving user expectations, adaptation strategies, and potential cognitive biases that may arise when interacting with generative models. By focusing on ChatGPT, this study ensures ecological validity while laying the groundwork for future research on broader AI ecosystems.

\subsection{Participant Recruitment}

This study recruited participants through the Amazon Mechanical Turk platform \footnote{https://www.mturk.com/} to examine the impact of functional fixedness on user interactions with LLM-enabled chat search. The project invited participants who are 18 or older and are native English speakers. Aligned with our study focus on functional fixedness, we particularly encouraged participation from crowd workers who are unfamiliar with or have no prior experience with ChatGPT or other LLM-enabled chatbots. To ensure data quality and engagement, participants were required to pass an attention check embedded within the intent-based expectation questionnaire, and their responses were screened for coherence and engagement. Those failing to meet these criteria were excluded from the final dataset. Each participant received fair compensation based on the estimated task completion time, adhering to ethical guidelines for online human-subject research and Institutional Review Board (IRB) requirements (University of Oklahoma IRB approval \#: 16867). The informed consent was obtained from all participants, ensuring that they were aware of the study's purpose and their data's confidentiality. Our final sample after data screening contains interaction and annotation data from 450 participants. Within the sample, 68.28\% of the participants self-identified as male, and 31.28\% of the participants self-identified as female. 0.22\% prefer not to disclose, and 0.22\% self-identified as non-binary. In terms of age group distribution, 7.27\% were between 18 to 24; 60.79\% of the participants were in the range of 25 to 34; 24.89\% of the sample were between 35 and 44; 4.19\% were in the range of 45 to 54; 2.42\% were between 55 to 64; and 0.44\% were 65 or older.


Each participant engaged in one of the six decision-making tasks designed to assess functional fixedness in LLM-based chat interactions, and the task assignments were rotated to ensure balanced data distribution (75 participants per task). These tasks required users to engage in multi-turn dialogues with the chat system, which allows for an analysis of how user expectations and functional fixedness affect users' behaviors in interactive sessions. The diverse range of participant backgrounds, prior experiences, and task types can help enhance the generalizability of the findings to different user groups interacting with LLM-enabled search systems under varying decision tasks. 

\subsection{Task Topics and Design}
To examine the impact of functional fixedness under varying motivating tasks, this study designed six human decision-making tasks involving varying domains (e.g., AI hiring, healthy diets, home insulation, carbon-neutral lifestyle). For task design, we defined and manipulated two key dimensions, aiming to expose participants to diverse types of decision-making tasks: 1) \textbf{Decision type}: \textit{choice-selection} or \textit{prioritization}; 2) \textbf{Decision complexity}: \textit{fixed dimension} or \textit{open-ended}. Regarding the decision type, \textit{choice-selection} refers to the tasks where participants were presented predefined options to compare and choose. \textit{prioritization} tasks require participants to evaluate and rank the offered options (or exploring open-ended options). With respect to the decision complexity, \textit{fixed dimension} tasks have predefined dimensions on which participants were asked to assess the available options. \textit{open-ended} tasks, however, require participants to explore and define reasonable dimensions to support the evaluation of options. Our tasks include:
\small  \begin{itemize}
   \item \textbf{Task 1}: Assess the impacts of using AI in hiring processes, considering various societal, cultural and ethical aspects. Should we use AI in hiring processes and why? [\textit{Choice selection - open ended}]
    \item \textbf{Task 2}: Investigate the balance between privacy and public safety in the use of surveillance cameras in urban areas, considering legal, ethical, and technological perspectives. Should we use surveillance cameras in urban areas and why? [\textit{Choice selection - dimension fixed}]
    \item \textbf{Task 3}: Compare the Keto, Vegan, and Mediterranean diets focusing on their nutritional profiles, suitability for various age groups, and effects on diabetes and heart health. Please rank the three diets and explain your decision. [\textit{Prioritization - dimension fixed}]
    \item \textbf{Task 4}: Evaluate the effectiveness of various digital tools and apps in managing and tracking symptoms of asthma, including emergency response features, long-term treatment adherence, and other relevant aspects. Please pick and rank your top three tools or apps and explain your decision. [\textit{Prioritization - open ended}]
    \item \textbf{Task 5}: Investigate and report on the effectiveness of three major home insulation materials (fiberglass, cellulose, and foam) in terms of energy efficiency, cost, and environmental impact. Please rank the three materials and explain your decision. [\textit{Prioritization - dimension fixed}]
    \item \textbf{Task 6}: Evaluate the feasibility and impact of adopting a carbon-neutral lifestyle, considering changes in diet, transportation, and energy use at home. Is it possible for all of us to adopt a carbon-neutral lifestyle? Why? [\textit{Choice selection - dimension fixed}]
\end{itemize} \normalsize

Before releasing and widely distributing the study link on crowdsourcing platform, we pilot tested the designed decision-making tasks and modified the task descriptions (e.g., adding dimensions, modifying options and requirements of comparison) according to pilot results to ensure that they can motivate multi-round interactions with chatbots and active changes in in-situ expectations and prompting strategies, rather than one-prompt or simple-intent short conversation. In addition, we selected task topics that are reasonably familiar to most participants and have practical implications and sought to avoid overly niche topics that are unlikely to be submitted to LLM-enabled chatbots by ordinary people in naturalistic settings. Participants responded to our topic familiarity question: \textit{To what extent were you familiar with the topic?}: 24.23\% of the participants reported that they were very familiar with the topic, while 3.74\% of the participants indicated that they were not familiar with the assigned topic at all. a majority of the participants fell into the range from slight familiar to moderately familiar (11.23\% slightly familiar; 22.47\% somewhat familiar; 38.33\% moderately familiar). This result indicates that our task and topic manipulation is successful in helping create the expected experimental condition for measuring functional fixedness, where we have participants under different levels of prior experience, and a majority of the participants are not very familiar with ChatGPT.

\subsection{Prior Experience with ChatGPT, Search Engines, and Virtual Assistants}
Due to the impact of functional fixedness, users' interactions with a unfamiliar tool or system (e.g., LLM-enabled chatbot) may be restricted or fixated by their prior experiences with ChatGPT and other similar AI chatbots, commercial search engines, and traditional virtual assistants that are not supported by generative AI. Thus, users' prior experiences with these systems may act as anchor points that shape users' intents, expectations, and interaction strategies. To facilitate the assessment on levels of functional fixedness, in pre-interaction survey, we collected data on how often do users use \textit{AI chatbot (e.g., ChatGPT)}, \textit{commercial search engines (e.g., Google, Microsoft Bing)}, and \textit{traditional virtual assistants (e.g., Alexa, Cortana)} respectively. Participants were asked to select one of the four frequency levels for each question: 1) \textit{Never}/This is going to be my first time; 2) \textit{Occasionally} (less than once a week); 3) \textit{Regularly} (once a week or more); 4) \textit{Frequently} (daily or almost daily). The descriptive statistics are reported in Table \ref{tab:usage_frequency_percentage}.

At the operationalization level, this study estimates functional fixedness based upon the correlation between users' conversational interactions and their prior experiences with the three types of systems. More specifically, at the cognitive level, this study investigates the extent to which participants' frequency of chatbot, search engine, and virtual assistants usage are correlated with their pre-chat intent-based expectations (i.e., what they expect ChatGPT can achieve). In addition, at the behavioral level, we examine how participants' chat interaction behaviors, such as prompt formulating strategies and response reading behavior, are associated with the frequencies of their prior system usage. Besides, from a dynamic perspective, this study explores how participants' experiences of prior system usage affect their prompt reformulation behaviors over the course of conversational sessions, and how unmet expectations might trigger unanticipated changes in prompting tactics, which can be potential opportunities to go beyond fixated approaches. 

\begin{table}[h]
    \centering
    \footnotesize
    \caption{Users' Prior Frequency of Usage: ChatGPT, Search Engine, Virtual Assistant.}
    \begin{tabular}{lccc}
        \hline
        \textbf{Usage Frequency} & \textbf{ChatGPT (\%)} & \textbf{Search Engine (\%)} & \textbf{Virtual Assistant (\%)} \\
        \hline
        Frequently (daily or almost daily) & 19.80 & 16.63 & 21.98 \\
        Regularly (once a week or more) & 22.01 & 41.02 & 43.96 \\
        Occasionally (less than once a week) & 30.46 & 40.58 & 28.57 \\
        Never/This is going to be my first time & 27.73 & 1.77 & 5.49 \\
        \hline
    \end{tabular}
    \label{tab:usage_frequency_percentage}
\end{table}

\subsection{User Intent Typology}
Before chat interactions actually occur, users' intent-based expectations, which in this work refers to the actions or tasks that a user expects a system can accomplish, could be fixated by their prior interaction experiences with the same or similar systems. A user's mental model and expected intent space may not be aligned with the actual functionalities of a new system, which is likely to lead to unsatisfied needs and ineffective interactions. To capture this cognitive level of functional fixedness, our study develops a \textit{user intent typology} based on existing typologies that describe users' general intents in search and chat interactions~\cite[e.g.,][]{rha2016exploring, bodonhelyi2024user} and participants' feedback and annotations from three rounds of pilot studies. Our typology building process started with integrating recently published preliminary chat intent typologies \cite[e.g.,][]{bodonhelyi2024user, wang2024user} and relevant search intents from IR research \cite[e.g.,][]{broder2002taxonomy, rha2016exploring, qu2018analyzing, mitsui2017predicting}. Then, through pilot testing, we identified new intents from participants' feedback, and also edited and removed the intents that do not make sense in chat interaction contexts or this study environment, or annotated as "never occurs" in more than 95\% of the pilot participants. Our goal is to build and apply a user intent typology that can properly capture the nuances and changes in user chat intents, achieves certain level of generalizability across different tasks and modalities and is topic-independent, and can facilitate the deconstruction of complex motivating tasks. Meanwhile, we seek to minimize the annotation effort of participants by prioritizing the intents that are most relevant and representative, frequently occur in users' chat search experiences, and best reflect the flexibility of LLMs compared to traditional search systems. Our finalized typology consists of following user intents and intent categories: 

\vspace{4pt}
\textbf{Category 1. \textit{Identify New Information}}

\small \begin{itemize}
    \item Identify an appropriate starting point to search or chat: For instance, find good query keywords, chat prompts/questions, or task description.
    \item Identify something more to search or learn: Explore a target topic or domain more broadly.
    \item Identify something new or unexpected: Explore a piece of  information, interest, advice that are completely new or unexpected; obtain inspiration or ideas for creative projects.
    \item Obtain Explanations: Search or ask for explanations or clarifications about phenomena, concepts, problems, or potential solutions.
\end{itemize} \normalsize

\vspace{2pt}
\textbf{Category 2. \textit{Find}}

\small \begin{itemize}
    \item Find a known item or site: Searching or asking for an item (e.g., a bag, book, name of a restaurant), website or information source that you were familiar with in advance.
    \item Find items sharing a named characteristic: Finding items or information with something in common.
    \item Find items without predefined criteria: Finding items that are potentially useful for a task, but which haven't been specified in advance
\end{itemize} \normalsize

\vspace{2pt}
\textbf{Category 3. \textit{Evaluate}}

\small \begin{itemize}
    \item Evaluate correctness of an item (e.g., retrieved information, generated chat response): Determine whether an item is factually correct.
    \item Evaluate usefulness of an item (e.g., retrieved information, generated chat response).
    \item Compare and pick best item(s) from all the useful ones.
    \item Evaluate users’ decision or judgment (e.g., analyzing the pros and cons, and risks associated with different options retrieved or generated).
\end{itemize} \normalsize

\vspace{2pt}
\textbf{Category 4. \textit{Advanced Information Processing}}

\small \begin{itemize}
    \item Content Creation: help in writing, proofreading, visually representing, or designing content.
    \item Synthesize and Summarize Online Information: Organize, synthesize, analyze, and/or summarize online information from the open Web.
    \item Synthesize and Summarize Previous Information: Organize, synthesize, analyze, and/or summarize the information retrieved or generated under your previous queries, questions or prompts within the current conversation/session with the system.
    \item Data Processing: Help in processing, analyzing, and visualizing data.
\end{itemize} \normalsize

\vspace{2pt}
\textbf{Category 5. \textit{Problem-Solving}}

\small \begin{itemize}
    \item Tutorial Requests: Step-by-step instructions or guidance.
    \item Technical Problems: Assistance with technical tasks or questions, coding, or problem-solving in a professional context.
    \item Societal and Ethical Queries: Exploring societal, cultural, or ethical topics.
\end{itemize} \normalsize

For \textit{each} intention, our annotation questionnaire includes two questions:

\vspace{4pt}
\textbf{Q1:} Based on your expectation, please classify the intent described above into one of the following categories.

\small \begin{itemize}
    \item ChatGPT can always fully fulfill the intention.
    \item ChatGPT may be able to fully fulfill the intention if/once an effective query/prompt is successfully formulated (e.g., after several rounds of query/prompt modifications).
    \item ChatGPT may be able to partially fulfill the intention if/once an effective query/prompt is successfully formulated.
    \item ChatGPT is unlikely to fulfill this intention at all. 
\end{itemize} \normalsize

\vspace{2pt}
\textbf{Q2:} In your prior interaction experiences, how often did you try to use ChatGPT to fulfill the intention described above?

\small \begin{itemize}
    \item Never or Rarely Used: less than once a month (very infrequent use, possibly for very specific or rare queries).
    \item Infrequently Used: about once a month (potentially for monthly tasks or when a specific need arises). 
    \item Moderately Used: engage with the system on a weekly basis. (represents a more habitual use, possibly for weekly tasks, regular inquiries, or as a part of a routine).
    \item Frequently Used: use the system several times a week. (as a regular part of the user’s “toolkit”). 
    \item Heavily Used: use the system daily or almost daily. It indicates a deep integration of the system into the user’s daily activities, using it as a primary tool for information, assistance, or interaction. 
\end{itemize} \normalsize

\begin{figure}[h]
    \centering
    \includegraphics[width=0.8\linewidth]{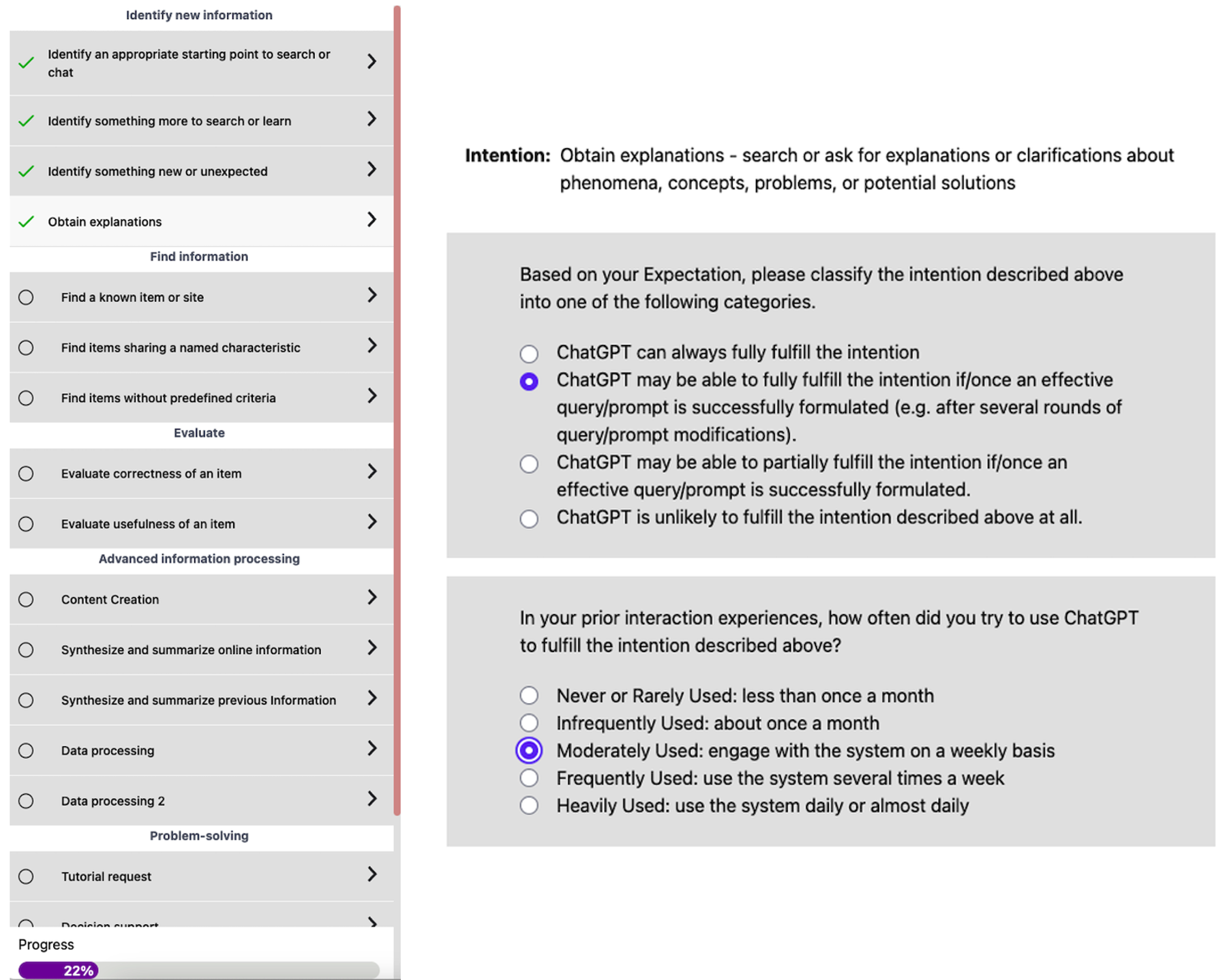}
    \caption{Pre-Chat Intent-based Expectation Annotation Interface.}
    \label{fig:intent}
\end{figure}

Figure~\ref{fig:intent} presents the intention-based expectation annotation page. The intents to be annotated are listed under different categories on the left sidebar. Once the participant completes the questionnaire under each intent, the system will assign a check mark to the intent and automatically move on to the next intent. Participants can also click on any of the listed intents to view or change their own answers. Note that as part of the data quality control measure, we randomly assign \textit{attention check} questions to different positions in the annotation questionnaire, aiming to check if the participant is actually paying attention to the annotation questions. 

\subsection{In-situ Annotation on Expectation Confirmation States}

Besides pre-interaction mental model and expectations, users' search interactions are also affected by their in-situ expectations regarding the system responses to their queries~\cite{wang2024understanding}. In particular, unmet in-situ expectations, while could lead to unsatisfactory conversation experience, may also trigger changes in prompting strategies and thus create opportunities to mitigate functional fixedness. After completing pre-chat demographic and intent-based expectation surveys, participants are invited to the chat interaction phase, which is up to 20 minutes. Participants are allowed to continue their response editing when the 20 minutes limit is passed, but can no longer issue a new prompt.

\begin{figure}[h]
    \centering
    \includegraphics[width=0.8\linewidth]{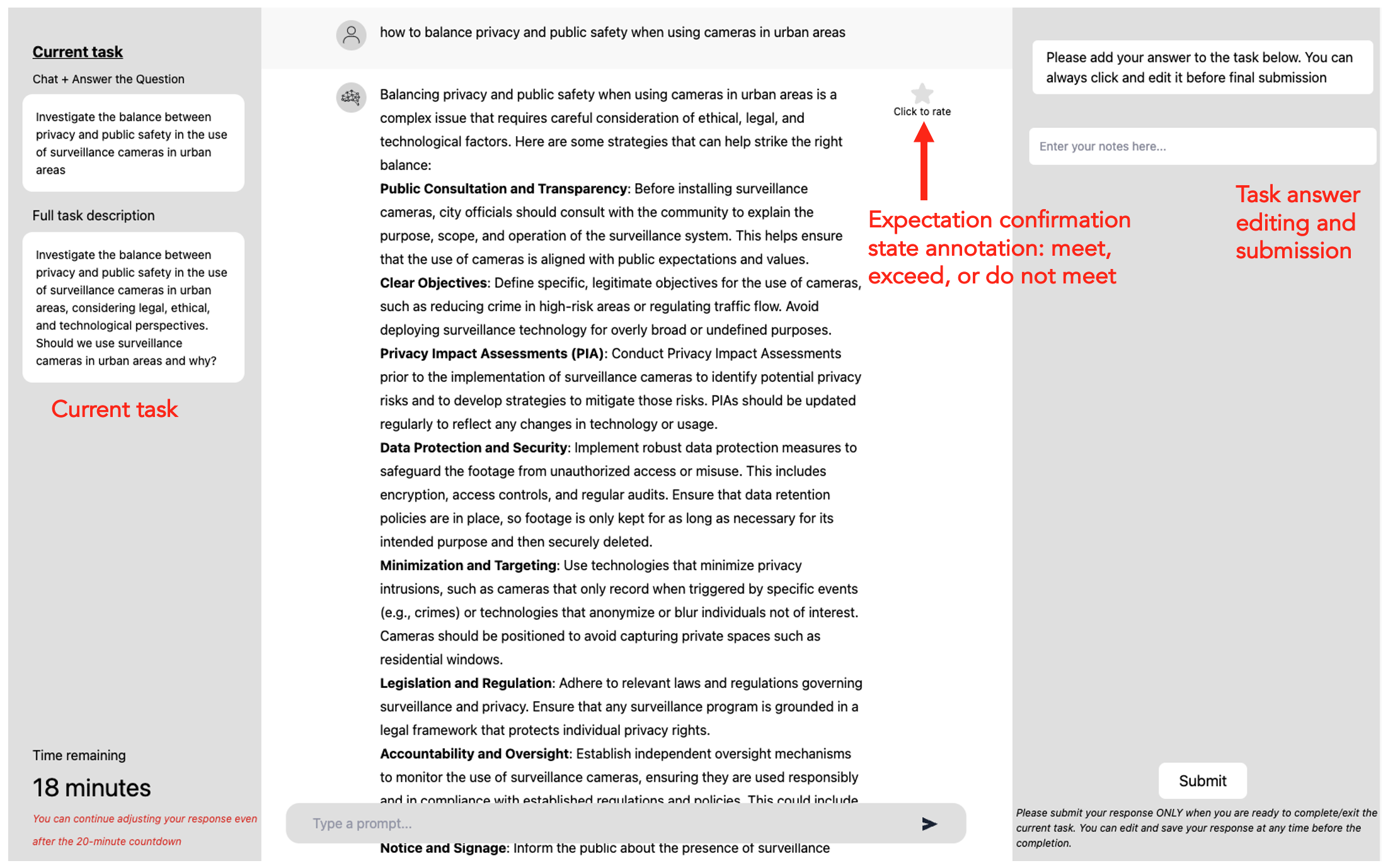}
    \caption{Chat Interaction Interface.}
    \label{fig:chat1}
\end{figure}

Figure~\ref{fig:chat1} presents the chat interaction interface. The assigned task is presented on the left sidebar, and the participant can write and edit their responses to the task question in the textbox on the right sidebar. During the chat interactions, participants are asked to annotate their expectation confirmation states under each prompt-response combination. The \textit{in-situ} annotation question (presented in a pop up window after clicking the star icon on the interface) is: 

\vspace{4pt}
\textbf{Q:} Does the system response under the current prompt meet my expectation?

\begin{itemize}
    \item The response \textit{exceeds} my expectation under the current prompt.
    \item The response \textit{meets} my expectation under the current prompt.
    \item The response \textit{does NOT meet} my expectation under the current prompt.
\end{itemize}

Each participant is required to submit at least four prompts to reasonably address the task question before being allowed to complete the session or submit their task response. Also, participants would receive a warning message and were not allowed to submit a new prompt if they did not complete the expectation confirmation state annotation for the previous prompt-response combination. 

\subsection{Crowdsourcing Experiment Procedure}

\begin{figure}[h]
    \centering
    \includegraphics[width=1\linewidth]{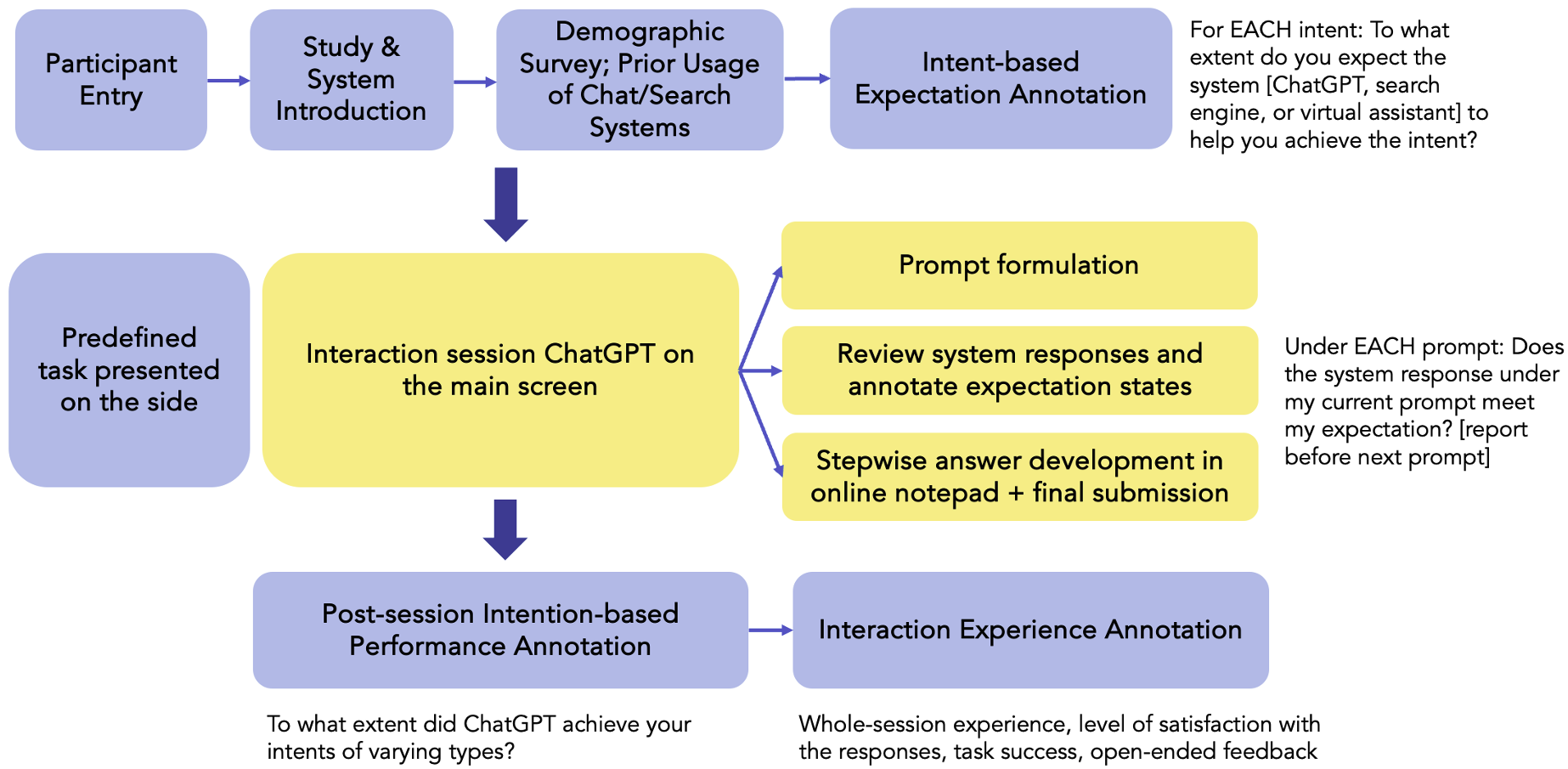}
    \caption{User Study Procedure.}
    \label{fig:procedure}
\end{figure}

Figure~\ref{fig:procedure} illustrates the crowdsourcing experiment procedure. After the chat session is completed, participants are invited to the post-session intent-based performance annotation, where they report the extent to which the chat system \textit{actually} fulfilled each intent. Then, the study is finished with an interaction experience survey, where participants answer questions regarding their whole-session chat experience, satisfaction level with system responses, and self-rated task success. 

\subsection{Pilot Testing and System Modifications}
To the best of our knowledge, this work is the first study that systematically examines functional fixedness in chat search contexts. Pilot testing and reporting are essential for understanding the design justification, limitation, and practical implication of a study, especially for user studies in IR and HCI research~\cite{liu2019interactive}. As a first step in this direction, in addition to the initial feasibility testing among the researchers, our work went through five rounds of pilot studies before the full-scale recruitment, which allowed us to test the feasibility of the experimental system, study procedure, and pre-defined decision-making tasks, and to modify these components when there were issues encountered in recruitment, users' understanding of the interface components and study flow, and overall data quality.  

During the first three rounds of pilot testing, we recruited 10 participants in each round, aiming to test the feasibility and effectiveness of study procedure. During these tests, we were able to identify and address several issues on both participant and system sides, such as participants' attention and effectiveness of attention checks, effectiveness of tasks in motivating rich chat interactions, appropriateness of user intent typology, and errors in the chat system. 

After addressing the major issues in study components with relatively small pilot testing sizes, In round-4 and round-5 pilot studies, we expanded to 30 participants and 60 participants respectively, aiming to test the modified study procedure and experimental chat system. In these two rounds of pilot studies, we finalized the attention check questions in both pre- and post-chat surveys; reduced the number of tasks from nine tasks to six tasks as three of the tasks (adapted from interactive IR research) were not effective or suitable in chat search contexts; adjusted our recruitment scope to MTurk works with 97\% approval rate \footnote{we tested both 95\% approval rate and 99\% approval rate with different amounts of HITs, and our results show that 97\% approval rate \textit{for this study} can achieve the balance between recruitment progress and data quality.} in prior tasks and MTurk Masters, which helped improve the data quality and also maintained the efficiency and speed in recruitment; removed some of the irrelevant and redundant intents from annotation (e.g., multimedia search and processing intent), and edited some of the annotation questions that are not clear enough to many participants; and also fixed some system bugs (e.g., the expectation confirmation state survey occasionally did not pop up properly for some prompts; a small set of user prompts were not clearly matched with corresponding tasks and system responses in dataset; for some users, the system did not stop them from submitting their responses when they issued less than four prompts). After completing five rounds of pilot tests, we employed the modified study flow, system, and survey tools, and collected data from 540 recruited crowd workers. The recruitment were restricted to English speaking countries in order to minimize the potential differences in task interpretations. We then checked the data quality, and filtered out the data from participants who either 1) completed all intent-based annotation questions within one minute, which, as we tested in pilot studies, was not possible for ensuring appropriate reading and annotation; or 2) failed the attention check in pre-chat or post-chat intent-based expectation annotation survey. After the data screening and quality control, we obtained and ran analyses on the study data from 450 participants. 

By presenting the pilot study progresses, results, and post-pilot adjustments, we hope to improve the transparency and replicability of our crowdsourcing study and also to better share the lessons we learned from the challenges, failures, and expected errors encountered at different levels to the researchers in this community.


\subsection{Data and Analysis}
The demographic results reveals a participant sample (N=450) with diverse representation across age groups, educational backgrounds, and occupational roles, which can help promote the broad generalizability of the findings. Among participants, 69\% held at least a bachelor’s degree, while 19\% reported high school level education, highlighting a well-educated cohort. Prior experience with AI-driven tools varied significantly: 37\% had never or only occasionally used ChatGPT, 33\% used it regularly, and 30\% engaged with it daily. In contrast, search engines remained a dominant tool, with 85\% using them at least weekly, while virtual assistants had a more polarized adoption, with 41\% rarely using them and 27\% relying on them frequently. These variations in prior experiences provide an appropriate empirical basis for examining functional fixedness, as those with extensive chatbot experience may formulate shorter, more task-specific prompts and interact more fluidly, while those with more prior interactions with search engines may use more structured, content-heavy prompts. The data underscores how prior exposure to different information-seeking and processing tools shapes users' intent-based expectations, interaction strategies, and whole-conversation experience. 

With respect to data analysis, to address \textbf{RQ1}, we conducted \textit{Spearman's rank correlation} analysis to examine the association between users' prior experiences (with ChatGPT and similar chatbots, commercial search engines, and virtual assistants) and their pre-chat intent-based expectations. This analysis explores the extent to which users' prior exposure to similar systems affect their expectations on what intent(s) ChatGPT is capable of fulfilling. Moving on to the behavioral level, to answer \textbf{RQ2}, we ran \textit{Kruskal-Wallis H tests} on the differences in users' chat interactions, especially prompting strategies, across different frequencies of previous ChatGPT/search engine/virtual assistant usage. With respect to the changes in prompting strategies, we examined the \textit{similarity of adjacent prompt pairs}, and explored if prior exposure to other systems may lead to smaller, incremental changes or major switches in prompting tactics in chat interactions. Besides, we also measured the frequency of different linguistic markers—including politeness markers, hedge words, discourse markers, deictic terms, and conversational fillers—within user-generated prompts. These linguistic features were selected based on their established role in conversational dynamics and cognitive adaptation processes. The frequency of each marker was calculated as the proportion of words in a user's prompt that matched predefined lists of relevant terms. Specifically, we computed the marker frequency for each user as:

\small \begin{equation}
    \text{Marker Frequency} = \frac{\text{Count of Marker Terms in User's Prompts}}{\text{Total Unique Words Used by the User}}
\end{equation} \normalsize

This approach provides a normalized measure of linguistic behavior, accounting for variations in users' overall verbosity and prompt writing styles. Table \ref{tab:linguistic_markers} in Appendix presents the representative terms for each type of linguistic markers we examined here. 

In addition, inspired by prior interactive IR research and findings on task impacts \cite[e.g.,][]{liu2021deconstructing, li2008faceted, choi2019effects, capra2018effects, kelly2015development}, this study investigated how participants' interaction strategies varied across decision tasks of varying types, as task may serve as a key moderating factor on functional fixedness. With respect to \textbf{RQ3}, we tested how a participant's current \textit{expectation confirmation state (i.e., meet, does not meet, exceed)} affect their prompting strategy and response reading behavior in the subsequent iteration. In addition, at the whole-session level, this study investigated how users' prompt-level expectation confirmation states affect their \textit{overall chat satisfaction} and \textit{self-reported task success}. 

\section{Results}
This section presents our analysis results and discusses how they address the proposed RQs. 





\subsection{RQ1: Correlation between Prior Experience and Pre-Chat Expectations}
The results in Table \ref{tab:intent_expectations} highlight the correlations between users' pre-chat expectations and their prior experiences with different digital systems. Specifically, the analysis compares the extent to which participants expect ChatGPT can fulfill a given intent, and also examining how this is associated with their prior frequencies of ChatGPT, search engine, and virtual assistant usage (e.g., \textit{does higher prior frequency of ChatGPT usage lead to higher user expectations with respect to certain intents}). To control for multiple comparisons, we applied the Benjamini-Hochberg correction, and only the correlations reported in the Table \ref{tab:intent_expectations} remained significant. We found that users with richer prior ChatGPT experiences were more likely to expect LLMs to handle societal and ethical queries ($\rho$ = 0.099, p < 0.05) and open-ended item-finding tasks ($\rho$ = 0.097, p < 0.05), which are relatively complex and cognitively demanding compared to traditional factual search intents. In contrast, users with more prior search engine experience demonstrated lower expectations for LLMs in summarization tasks, as indicated by negative correlations for synthesizing online ($\rho$ = -0.099, p < 0.05) and previous-session information ($\rho$ = -0.102, p < 0.05) with prior frequency of search engine usage. Similarly, users with richer experience on using virtual assistant support exhibited lower expectations for LLMs in tasks such as tutorial requests ($\rho$ = -0.117, p < 0.05) and decision support ($\rho$ = -0.101, p < 0.05). These findings suggest that users' pre-existing mental models and expectations, shaped by their experience with specific systems, play a crucial role in determining what they believe LLM-enabled chat systems can accomplish. Also, prior exposures to different systems may act differently in shaping users' intents and expectations. 

\begin{table}[h]
    \centering
    \footnotesize
    \caption{Correlation of Intent-based Expectations and Frequency of Previous System Usage}
    \begin{tabular}{p{6cm} c c}
        \toprule
        \textbf{Users' Intent in Chat Interactions} & \textbf{System Type} & \textbf{Spearman’s $\rho$} \\
        \midrule
        Societal and ethical queries & ChatGPT & 0.099* \\
        Find items without predefined criteria & ChatGPT & 0.097* \\
        \midrule
        Synthesize and summarize online information & Search Engine & -0.099* \\
        Synthesize and summarize previous information & Search Engine & -0.102* \\
        \midrule
        Synthesize and summarize previous information & Virtual Assistant & -0.121* \\
        Tutorial request & Virtual Assistant & -0.117* \\
        Finding items sharing a named characteristic & Virtual Assistant & -0.114* \\
        Decision support & Virtual Assistant & -0.101* \\
        \bottomrule
    \end{tabular}
    \label{tab:intent_expectations}
\end{table}

These results address \textbf{RQ1} by demonstrating that functional fixedness manifests at the intent-based expectation level, as users transfer assumptions regarding system functionalities from previously used tools to LLM-enabled chat interactions. Participants with rich chatbot experiences, having engaged in generative dialogues, anticipate more flexible, open-ended capabilities, whereas search engine users, accustomed to retrieving pre-existing content, do not expect LLMs to effectively synthesize information beyond a static dataset. Likewise, participants using virtual assistants frequently, familiar with task-specific directives, exhibit skepticism toward LLMs' ability to support complex decision-making. These findings underscore the need for system designs that help users recalibrate their expectations, potentially through adaptive affordance signaling, to mitigate functional fixedness and encourage active exploration and utilization of generative affordances.



\subsection{RQ2: Correlation between Prior Experience and Chat Interactions}
\subsubsection{Functional Fixedness in Chat Interactions}
RQ1 examines functional fixedness at a \textit{cognitive level} by testing the correlation between prior interaction experiences and the intents that users expect ChatGPT to fulfill. Moving on to \textit{behavioral level}, the results presented in Table \ref{tab:behavior_chat}, Table \ref{tab:behavior_search}, and Table \ref{tab:behavior_virtual} reveal distinct behavioral patterns in how users interact with LLM-enabled chat systems based on their prior experiences with ChatGPT, search engines, and virtual assistants. Table \ref{tab:behavior_chat} shows that users with frequent ChatGPT experience tended to issue shorter prompts (median length: 45.4 characters) with fewer unique words (median: 7.55) and formulated prompts more quickly (median: 9.6 seconds). They also exhibited lower between-prompt similarity ($\rho$ = 0.106, p < 0.01), indicating a more flexible and exploratory approach to prompt formulation. This suggests that users with more ChatGPT usage experiences are more accustomed to engaging in fluid, iterative prompting strategies, likely reflecting their familiarity with LLM affordances. In contrast, Table \ref{tab:behavior_search} highlights that users with extensive search engine experience took significantly longer to construct each prompt (median: 13.75 seconds), and their prompts were both longer (median: 60.3 characters) and contained more unique words (median: 9.8). Additionally, users with more search experiences demonstrated higher between-prompt similarity, suggesting a tendency to refine prompts in an incremental, structured manner rather than experimenting with diverse phrasing or novel prompting tactics.

\begin{table}
\centering
\footnotesize 
\caption{Behavioral Differences in Chat Search Across Different Frequencies of Previous ChatGPT Usage}
\label{tab:chat_behavior}

\begin{tabular}{@{}p{6cm}ccc@{}}
\toprule
\textbf{Behavioral Measure} & \textbf{Never/Occasionally} & \textbf{Regularly} & \textbf{Frequently} \\ 
\midrule
Prompt formulation time & 12 (20.44) & 7.2 (16.21) & 9.625 (13.75) \\
\textbf{Prompt length (in characters)**} & 59.4 (71.5) & 35.6 (53.58) & 45.375 (56.57) \\
\textbf{Number of unique words in prompt**} & 9.75 (10) & 6 (7.25) & 7.55 (6.71) \\
\textbf{Number of prompts within session**} & 5 (0) & 4 (1) & 4 (1) \\
\textbf{Average prompt rating within session*} & 2.4 (0.75) & 2.5 (0.8) & 2.6 (0.76) \\
\textbf{Average between-prompt similarity**} & 0.155 (0.20) & 0.114 (0.14) & 0.106 (0.12) \\
\textbf{Response reading time (sec)*} & 56.5 (79.7) & 41.0 (45) & 51.275 (37.16) \\
\textbf{Response length/reading time (per sec)*} & 52.785 (42.41) & 56.937 (42.84) & 61.753 (38.50) \\
\bottomrule
\label{tab:behavior_chat}
\end{tabular}

\vspace{2mm}
\begin{flushleft}
    \small
    \hspace{1cm}
    \textit{Note: The table presents the results from Kruskal-Wallis tests, with median value (and IQR in the parentheses) under each frequency level. *: $p<0.05$. **: $p<0.01$. Frequency levels: \textbf{Never} = This is going to be my first time using AI Chatbot; \textbf{Occasionally} = less than once a week; \textbf{Regularly} = once a week or more; \textbf{Frequently} = daily or almost daily.}
\end{flushleft}

\end{table}

\begin{table}
\centering
\footnotesize 
\caption{Behavioral Differences in Chat Search Across Different Frequencies of Previous Search Engine Usage}
\label{tab:search_engine_behavior}
\begin{tabular}{@{}p{6cm}ccc@{}}
\toprule
\textbf{Behavioral Measure} & \textbf{Never/Occasionally} & \textbf{Regularly} & \textbf{Frequently} \\ 
\midrule
\textbf{Prompt formulation time**} & 6.85 (14.94) & 6.75 (13.82) & 13.75 (16.83) \\
\textbf{Prompt length (in characters)**} & 22.65 (46.63) & 37.4 (66.07) & 60.25 (55.2) \\
\textbf{Number of unique words in prompt**} & 3.675 (7.06) & 6.25 (9.2) & 9.8 (7.25) \\
Number of prompts within session & 4.0 (0) & 4.0 (1) & 4.0 (1) \\
Average prompt rating within session & 2.25 (0.75) & 2.5 (0.72) & 2.5 (0.55) \\
Average between-prompt similarity & 0.145 (0.30) & 0.113 (0.14) & 0.126 (0.15) \\
\textbf{Response reading time (sec)**} & 36.125 (58.29) & 41.5 (42.96) & 61.5 (59.75) \\
\textbf{Response length/reading time (per sec)*} & 51.173 (47.69) & 61.342 (47.27) & 56.157 (34.43) \\
\bottomrule
\label{tab:behavior_search}
\end{tabular}
\end{table}


Table \ref{tab:behavior_virtual} further illustrates that users with more virtual assistant interaction experiences exhibited a different pattern, often issuing structured and directive-style prompts. While their median prompt length (49.2 characters) and number of unique words (8.3) were comparable to other groups, they interacted with fewer prompts per session (median: 4.0) and showed a slower rate of adaptation. Unlike ChatGPT users, who adjusted their strategies dynamically, virtual assistant users appeared to engage in more rigid, command-like interactions. This suggests that their expectations, shaped by prior experience with AI assistants designed for task execution rather than open-ended dialogue, constrained their ability to leverage the conversational nature of LLMs. Additionally, across all groups, response reading time varied significantly, with search engine users spending the longest time per response (median: 61.5 seconds), followed by virtual assistant users (median: 52.8 seconds), and ChatGPT users (median: 51.3 seconds). This pattern suggests that users accustomed to search engines may engage in more deliberate reading and verification behaviors, whereas frequent ChatGPT users might be more confident in interpreting chat responses quickly.

\begin{table}[htbp]
    \centering
    \caption{Behavioral Differences in Chat Search Across Different Frequencies of Previous Virtual Assistant Usage}
    \label{tab:virtual_assistant_behavior}
    \footnotesize

    \begin{tabular}{@{}p{0.5\textwidth}ccc@{}}
        \toprule
        \textbf{Behavioral Measure} & \textbf{Never/Occasionally} & \textbf{Regularly} & \textbf{Frequently} \\ 
        \midrule
        \textbf{Prompt formulation time**} & 13.775 (19.89) & 7.708 (13.08) & 10.325 (15.09) \\
        Prompt length (in characters) & 53.325 (71.63) & 37.638 (69.08) & 49.167 (52.5) \\
        Number of unique words in prompt & 8.325 (10.25) & 6.325 (9.31) & 8.375 (6.33) \\
        \textbf{Number of prompts within session*} & 5.0 (1.0) & 4.0 (1.0) & 4.0 (1.0) \\
        \textbf{Average prompt rating within session*} & 2.333 (0.67) & 2.5 (0.70) & 2.5 (0.75) \\
        Average between-prompt similarity & 0.140 (0.19) & 0.115 (0.15) & 0.116 (0.12) \\
        Response reading time (sec) & 51.486 (68.13) & 45.571 (43.88) & 52.775 (39.09) \\
        \textbf{Response length/reading time (per sec)*} & 52.486 (39.40) & 59.545 (41.82) & 59.640 (38.60) \\
        \bottomrule
        \label{tab:behavior_virtual}
    \end{tabular}
\end{table}

These findings help answer RQ2 by showing how functional fixedness manifests in users’ prompting strategies and adaptation patterns during LLM interactions. Users' prior experiences with different digital systems influence both their initial prompt formulation behaviors and their ability to adjust strategies within a session. users who used ChatGPT frequently, having already internalized the system’s affordances, engage in more rapid, flexible prompting. In contrast, users who used search engines frequently, tend to be accustomed to precise keyword-based retrieval and take a more methodical approach, refining prompts with greater deliberation but limited structural shifts. participants with rich virtual assistant experiences, on the other hand, maintain directive prompting strategies and treat LLMs more like command-based systems rather than interactive dialogue partners. These behavioral tendencies suggest that prior system exposure constrains users' willingness or ability to explore LLM capabilities beyond familiar interaction paradigms. 

\subsubsection{Functional Fixedness in Prompts}
\begin{figure}[h]
    \centering
    \includegraphics[width=0.9\linewidth]{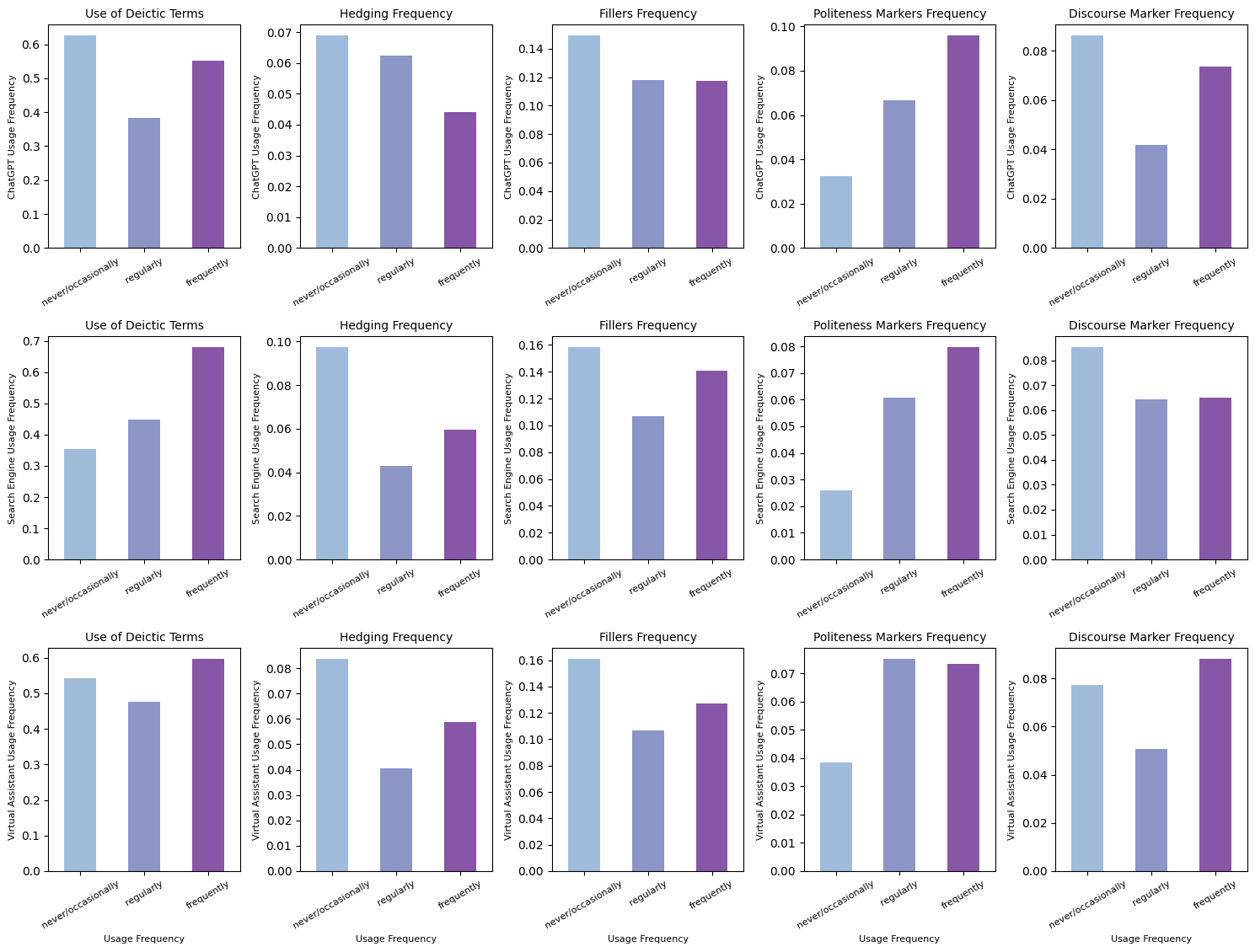}
    \caption{Behavioral Variable Frequencies by ChatGPT, Search Engine, and Virtual Assistant Usage}
    \label{fig:prompt}
\end{figure}

Going further on examining prompting strategies, the results in Figure \ref{fig:prompt} reveal distinct patterns in linguistic behaviors based on users' prior experience with ChatGPT, search engines, and virtual assistants. Participants who have more prior experience with ChatGPT used fewer deictic terms (e.g., “this,” “that,” “here”) compared to non-frequent or never-use participants, suggesting that they framed their prompts in a more explicit and self-contained manner rather than relying on conversational context. This could indicate that frequent ChatGPT users, despite their familiarity with the system, do not always assume it retains contextual memory across turns. Additionally, they used more politeness markers (e.g., “please,” “could you”) than non-frequent users, which suggests that they may have grown accustomed to treating the system more as a conversational partner. Conversely, users with less ChatGPT experiences showed higher use of both deictic terms, indicating a more exploratory engagement style. This suggests that functional fixedness among frequent ChatGPT users may manifest not only in how they structure prompts but also in how they linguistically frame their interactions, potentially limiting their willingness to engage in a more flexible conversational exchange.  

With respect to search, users with richer prior search engine experiences showed high reliance on deictic terms, indicating that they are more familiar with dialogue style interactions than non-frequent users. However, compared to frequent ChatGPT users, users who actively engage in traditional search activities use less hedge words and discourse markers, suggesting that they tend not to treat the search conversation as a natural conversation with a human-like agent. Regarding virtual assistant usage, those with frequent prior experience exhibited the lowest frequency of hedge terms and conversation fillers, with their prompts being more directive and command-like. This aligns with their familiar interaction style, where virtual assistants are typically used for issuing commands rather than engaging in multi-turn conversations. These findings highlight that functional fixedness in LLM interactions extends beyond prompt structure to linguistic choices, with frequent ChatGPT users favoring direct and detached phrasing, users with richer search experiences maintaining a rigid, independent-query approach, and users with frequent virtual assistant usage adhering to a task-execution style. These ingrained behavioral patterns may restrict how users explore LLMs and suggest that subtle interventions, such as contextual affordance cues or adaptive feedback, could help encourage more flexible, conversational engagement. 

\begin{figure}[h]
    \centering
    \begin{subfigure}[b]{1\textwidth}
         \centering
         \includegraphics[width=\textwidth]{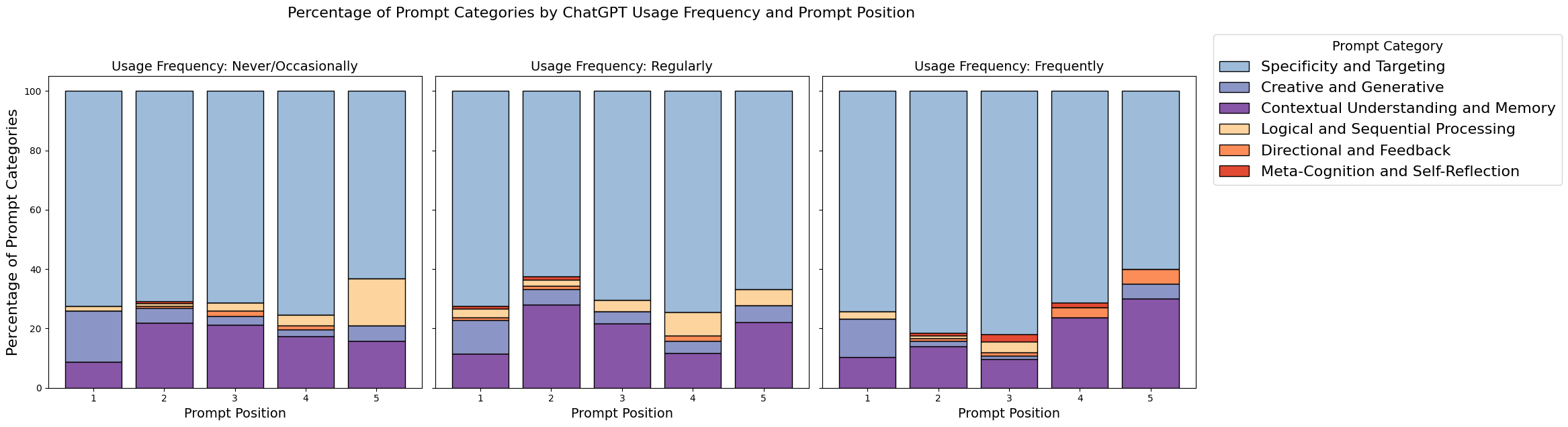}
         \label{fig:freq_chat}
    \end{subfigure}
    
    \vspace{1em} 

    \begin{subfigure}[b]{1\textwidth}
         \centering
         \includegraphics[width=\textwidth]{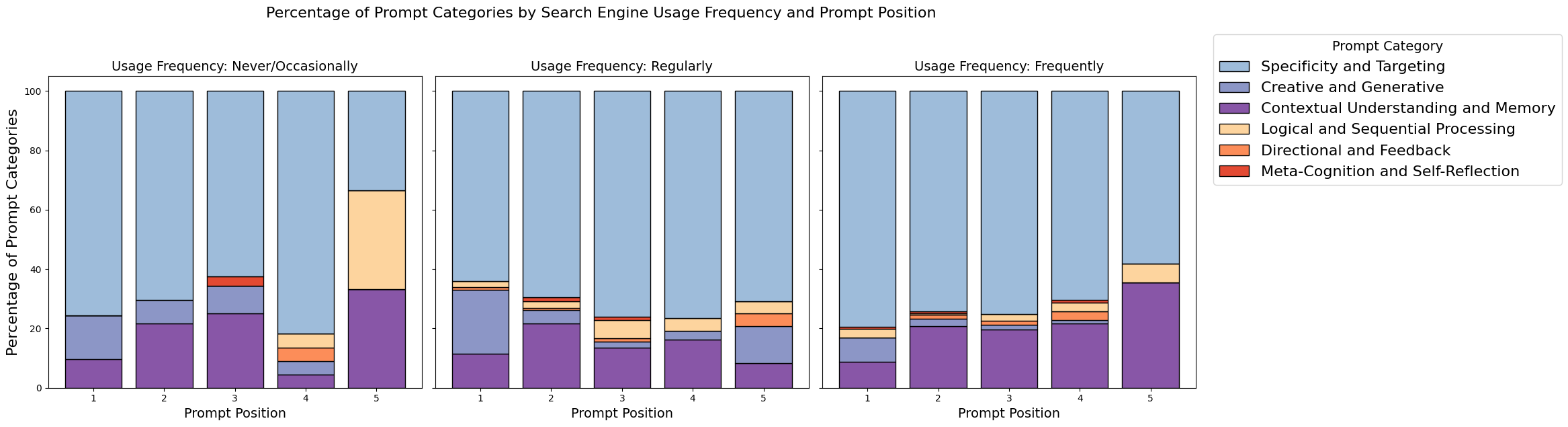}
         \label{fig:freq_search}
    \end{subfigure}
    
    \vspace{1em} 

    \begin{subfigure}[b]{1\textwidth}
         \centering
         \includegraphics[width=\textwidth]{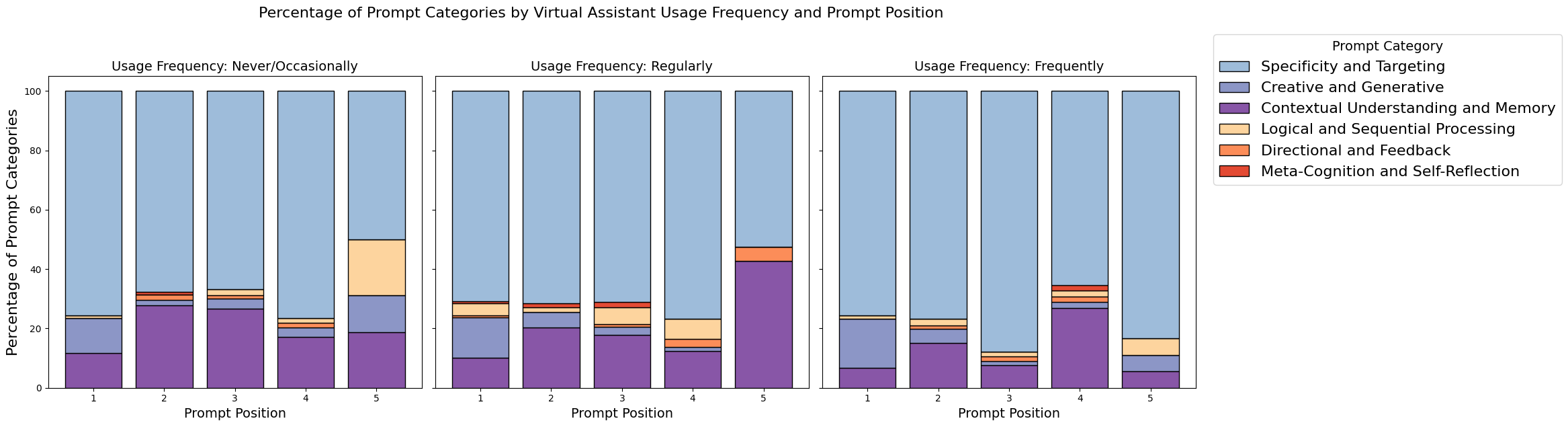}
         \label{fig:freq_virtual}
    \end{subfigure}
    
    \setlength{\belowcaptionskip}{0pt} 
    \caption{Percentage of Prompt Categories by ChatGPT, Search Engine and Virtual Assistant Usage Frequency and Prompt Position.}
    \label{fig:threefigs}
\end{figure}

To further examine how users' prompting strategies vary across different prompt positions and how this variation is related to previous interaction experiences, this study adopts the prompt typology developed and tested by~\cite{fagbohun2024empirical} to categorize crowdworker-issued prompts. Based on the prompt typology, we classified participants' prompts into five categories: Contextual understanding and memory, creative and generative, directional and feedback, logical and sequential processing, meta-cognition and self-reflection, and specificity and targeting. The classification is completed automatically through the GPT-4 model with the prompt template in the Appendix, where we also provided example prompts to illustrate each category (see Table \ref{tab:exampleprompt}). According to the results reported in Figure~\ref{fig:threefigs}, a majority of users' prompts fall under the specificity and targeting category, including the requests that improve the precision of LLMs and encourage more specific goal-oriented responses. However, users who interact with chatbots or search engines frequently tend to issue more contextual understanding and memory prompts as their conversations with the system proceed. This category covers prompts that recall and reference previous interactions to deliver seamless chat experiences and enhance coherence in extended conversations. For users who have a rich interaction experience with virtual assistants, this trend is almost reversed: Users tend to issue more specificity and targeting prompts as the chat interactions continue. Users with fewer previous virtual assistant experiences, however, often issue more prompts under the contextual understanding and memory category and the creative and generative category. Overall, the results show that users with rich prior experience in chat and search are more active in maintaining and enhancing the contextual coherence in multiprompt conversations. Users with fewer chat and search experiences, however, tend to issue more exploratory and sequential processing prompts (i.e., breaking down complex reasoning tasks) across different positions of the conversation. This finding indicates that while previous chat and search experience may contribute to users' skills in maintaining the coherence and relevance of chat sessions, it could also reduce the motivation for exploratory and creative prompting and limit the potential for information serendipity~\cite{fu2024art}. 

The results in Table \ref{tab:simprompt_experience} provide insights into how users with different levels of prior experience with ChatGPT, search engines, and virtual assistants exhibit varying levels of prompt similarity over the course of their interactions. To measure the connections between adjacent prompts, we calculated \textit{Cosine Similarity} scores based on the embeddings of the prompts. According to the results, participants with little to no prior experience showed significantly higher prompt similarity across consecutive turns, indicating that they tended to minimally modify their prompts rather than exploring alternative phrasing. In contrast, frequent ChatGPT users exhibited lower prompt similarity, suggesting that they were more likely to experiment with different ways of prompting and adapt their strategies. This pattern aligns with the idea that users with more prior experiences may have a better understanding of LLM capabilities and engage in more exploratory interactions, whereas inexperienced users struggle to move beyond their initial mental models. Additionally, average similarity scores across all prompts show that non-frequent ChatGPT users maintained more rigid and incremental modifications, further reinforcing the notion that prior familiarity with ChatGPT reduces functional fixedness by encouraging more flexible prompting behavior. 

\begin{table}
\footnotesize
\caption{}
\label{tab:similarity-score}

\begin{subtable}{\textwidth}
    \centering
    \caption*{Similarity Score of Prompts Across Different Frequencies of Previous ChatGPT Usage}
    \begin{tabular}{@{}p{0.4\textwidth}ccc@{}}
    \toprule
    \textbf{Prompt Positions} & \textbf{Never/Occasionally} & \textbf{Regularly} & \textbf{Frequently} \\ \midrule
    \textbf{Prompt 2 vs Prompt 1**} & 0.38 (0.56) & 0.23 (0.48) & 0.25 (0.49) \\
    \textbf{Prompt 3 vs Prompt 2**} & 0.37 (0.51) & 0.19 (0.31) & 0.19 (0.40) \\
    \textbf{Prompt 4 vs Prompt 3**} & 0.33 (0.52) & 0.18 (0.26) & 0.23 (0.37) \\
    \textbf{Prompt 5 vs Prompt 4**} & 0.25 (0.43) & 0.09 (0.12) & 0.11 (0.24) \\
    \bottomrule
    \end{tabular}
\end{subtable}

\vspace{3mm}

\begin{subtable}{\textwidth}
    \centering
    \caption*{\begin{minipage}{0.8\textwidth}
    Average Similarity Score of Each Prompt with Previous Prompts (by Frequencies of ChatGPT Usage)
    \end{minipage}}
    \begin{tabular}{@{}p{0.4\textwidth}ccc@{}}
    \toprule
    \textbf{Prompt Positions} & \textbf{Never/Occasionally} & \textbf{Regularly} & \textbf{Frequently} \\ \midrule
    \textbf{Prompt 2*} & 0.38 (0.56) & 0.23 (0.48) & 0.25 (0.49) \\
    \textbf{Prompt 3**} & 0.39 (0.45) & 0.27 (0.24) & 0.33 (0.32) \\
    \textbf{Prompt 4**} & 0.37 (0.38) & 0.25 (0.25) & 0.27 (0.26) \\
    Prompt 5 & 0.30 (0.34) & 0.22 (0.24) & 0.28 (0.20) \\
    \bottomrule
    \end{tabular}
\end{subtable}

\vspace{3mm}

\begin{subtable}{\textwidth}
    \centering
    \caption*{Similarity Score of Prompts Across Different Frequencies of Previous Search Engine Usage}
    \begin{tabular}{@{}p{0.4\textwidth}ccc@{}}
    \toprule
    \textbf{Prompt Positions} & \textbf{Never/Occasionally} & \textbf{Regularly} & \textbf{Frequently} \\ \midrule
    Prompt 2 vs Prompt 1 & 0.26 (0.34) & 0.23 (0.49) & 0.40 (0.56) \\
    Prompt 3 vs Prompt 2 & 0.17 (0.28) & 0.19 (0.35) & 0.36 (0.56) \\
    \textbf{Prompt 4 vs Prompt 3*} & 0.15 (0.24) & 0.20 (0.35) & 0.30 (0.43) \\
    \textbf{Prompt 5 vs Prompt 4**} & 0.11 (0.23) & 0.08 (0.10) & 0.18 (0.33) \\
    \bottomrule
    \end{tabular}
\end{subtable}

\vspace{3mm}

\begin{subtable}{\textwidth}
    \centering
    \caption*{\begin{minipage}{0.8\textwidth}
Average Similarity Score of Each Prompt With Previous Prompts (by Frequencies of Search Engine Usage)
\end{minipage}}
    \begin{tabular}{@{}p{0.4\textwidth}ccc@{}}
    \toprule
    \textbf{Prompt Positions} & \textbf{Never/Occasionally} & \textbf{Regularly} & \textbf{Frequently} \\ \midrule
    Prompt 2 & 0.26 (0.34) & 0.23 (0.49) & 0.40 (0.56) \\
    \textbf{Prompt 3**} & 0.20 (0.33) & 0.29 (0.27) & 0.39 (0.39) \\
    \textbf{Prompt 4*} & 0.22 (0.26) & 0.29 (0.26) & 0.30 (0.32) \\
    Prompt 5 & 0.24 (0.27) & 0.22 (0.21) & 0.32 (0.27) \\
    \bottomrule
    \end{tabular}
\end{subtable}
\begin{flushleft}
    \small
    \hspace{1cm}
    \textit{Note: The table presents the results from Kruskal-Wallis tests, with median value (and IQR in the parentheses) under each frequency level. *: $p<0.05$. **: $p<0.01$. Note that in chat sessions, participants have to run at least four prompts before final response submission/end the task. }
\end{flushleft}
\label{tab:simprompt_experience}
\end{table}


With respect to prior search engine experiences, the trend was reversed—frequent search engine users exhibited higher prompt similarity than their non-frequent counterparts, indicating that they tended to refine their queries in smaller, iterative steps rather than engaging in significant reformulation. This suggests that users using search engine more actively tend to carry over their keyword-based query refinement strategies to LLM-based interactions, treating each prompt as a structured query that requires minor modifications rather than fundamental restructuring. Similarly, frequent virtual assistant users also demonstrated higher prompt similarity, reinforcing their tendency to issue directive, command-like queries that follow a fixed format. These findings provide further evidence that functional fixedness may constrain users’ motivation or ability to adjust their prompting behavior when transitioning from traditional search and virtual assistants to LLM-based interactions. The results suggest that system interventions, such as dynamic prompt suggestions, that encourage linguistic variation, could help mitigate these rigid prompting patterns and facilitate more adaptive, exploratory chat interactions.

\begin{table}[htbp]
    \centering
    \caption{}
    \label{tab:prompt-similarity}
    \footnotesize

    \begin{subtable}{\textwidth}
        \centering
        \caption*{\begin{minipage}{0.8\textwidth}
        Prompt Similarity with Task Descriptions Across Frequencies of Previous ChatGPT Usage
        \end{minipage}}
        \begin{tabular}{@{}p{0.4\textwidth}ccc@{}}
            \toprule
            \textbf{Prompt Positions} & \textbf{Never/Occasionally} & \textbf{Regularly} & \textbf{Frequently} \\ 
            \midrule
            \textbf{Prompt 1**} & 0.72 (0.84) & 0.59 (0.59) & 0.36 (0.76) \\ 
            \textbf{Prompt 2**} & 0.51 (0.67) & 0.16 (0.48) & 0.34 (0.62) \\ 
            \textbf{Prompt 3**} & 0.48 (0.52) & 0.10 (0.39) & 0.29 (0.52) \\ 
            \textbf{Prompt 4**} & 0.43 (0.59) & 0.10 (0.39) & 0.20 (0.44) \\ 
            \bottomrule
        \end{tabular}
    \end{subtable}

    \vspace{3mm} 

    \begin{subtable}{\textwidth}
        \centering
        \caption*{\begin{minipage}{0.8\textwidth}
        Prompt Similarity with Task Descriptions Across Frequencies of Previous Search Engine Usage
        \end{minipage}}
        \begin{tabular}{@{}p{0.4\textwidth}ccc@{}}
            \toprule
            \textbf{Prompt Positions} & \textbf{Never/Occasionally} & \textbf{Regularly} & \textbf{Frequently} \\ 
            \midrule
            \textbf{Prompt 1**} & 0.10 (0.59) & 0.59 (0.59) & 0.36 (0.76) \\ 
            \textbf{Prompt 2**} & 0.16 (0.49) & 0.22 (0.53) & 0.50 (0.59) \\ 
            \textbf{Prompt 3**} & 0.13 (0.47) & 0.15 (0.49) & 0.43 (0.53) \\ 
            \textbf{Prompt 4**} & 0.10 (0.30) & 0.17 (0.44) & 0.30 (0.53) \\ 
            \bottomrule
        \end{tabular}
    \end{subtable}
    \label{tab:simprompt_task}
\end{table}


\subsubsection{Functional Fixedness and Task Effects}
The results in Table \ref{tab:simprompt_task} illustrate the extent to which users’ prompts remained similar to the original task description across different levels of prior experience with ChatGPT and search engines. Participants with little to no prior experience on ChatGPT demonstrated significantly higher prompt similarity to the task description in their initial and subsequent turns, indicating that they relied heavily on the given instructions rather than reinterpreting the task in their own words. In contrast, frequent ChatGPT users exhibited lower prompt similarity, suggesting that they were more comfortable modifying their prompts beyond the initial task framing. This pattern implies that non-frequent ChatGPT users may struggle to adapt their prompts dynamically, potentially due to a lack of familiarity with LLM affordances. participants using ChatGPT frequently, on the other hand, appeared more willing to restructure their queries in a way that better aligns with their evolving needs and system capabilities, which indicates their reduced fixedness compared to infrequent users.

Regarding search experiences, participants with richer search engine experiences exhibited higher prompt similarity with the task description compared to non-frequent users, indicating that they tended to adhere more closely to the original wording. This suggests that experienced search engine users, accustomed to precise and structured query formulation, were less likely to experiment with alternative phrasing, potentially due to their reliance on traditional search heuristics. In contrast, non-frequent search engine users demonstrated lower prompt similarity, suggesting greater flexibility in adapting their phrasing. With respect to levels of prior experience with virtual assistants, we did not observe cross-group differences at any prompt positions. 

These results further support RQ2 by demonstrating how functional fixedness manifests in users' adherence to given instructions, with frequent search engine and virtual assistant users displaying more rigid, instruction-bound behavior, while frequent ChatGPT users exhibited more flexible and adaptive prompting. This suggests that system interventions, such as proactive reformulation suggestions or interface elements encouraging paraphrasing, could help users break free from rigid query patterns and explore more effective ways to engage with LLM-based chat systems. 

\definecolor{lightpink}{RGB}{255, 228, 225}
\definecolor{lightblue}{RGB}{173, 216, 230}
\begingroup
    \fontsize{15pt}{12pt}\selectfont
\begin{table}
    \centering
    \footnotesize
    \caption{ }
    \label{tab:ai_task_types}

    \begin{subtable}{\textwidth}
        \centering
        \caption*{Behavioral Differences Across Frequencies of ChatGPT Usage Under Varying Task Types}
        \begin{tabular}{p{4cm}|c|c|c|c}
            \hline
            \textbf{} & \textbf{Choice-fixed} & \textbf{Prioritize-fixed} & \textbf{Choice-open} & \textbf{Prioritize-open} \\ 
            \hline
            Prompt formulation time & & \cellcolor{lightpink} - & \cellcolor{lightpink} - & \cellcolor{lightpink} - \\ 
            \hline
            Prompt length & & \cellcolor{lightpink} - & \cellcolor{lightpink} - &  \\ 
            \hline
            Number of unique words & & \cellcolor{lightpink} - & \cellcolor{lightpink} - &  \\ 
            \hline
            Number of prompts & &  & & \\ 
            \hline
            Average prompt rating & & \cellcolor{lightblue} + & & \\ 
            \hline
            Average between-prompt similarity & & \cellcolor{lightpink} - &  \cellcolor{lightpink} - & \\ 
            \hline
            Response reading time & & \cellcolor{lightpink} - & & \\ 
            \hline
            Response length/reading time & & & & \\ 
            \hline
        \end{tabular}
    \end{subtable}
    \vspace{1mm}
    \footnotesize
    \textit{Note:} \textbf{Light blue:} positive correlation. 
    \textbf{Light pink}: negative correlation. 
    \textbf{No color}: no significant connection.

    \begin{subtable}{\textwidth}
        \centering
        \caption*{Behavioral Differences Across Frequencies of Search Engine Usage Under Varying Task Types}
        \begin{tabular}{p{4cm}|c|c|c|c}
            \hline
            \textbf{} & \textbf{Choice-fixed} & \textbf{Prioritize-fixed} & \textbf{Choice-open} & \textbf{Prioritize-open} \\ 
            \hline
            Prompt formulation time & \cellcolor{lightblue} + & & \cellcolor{lightblue} + & \\ 
            \hline
            Prompt length & \cellcolor{lightblue} + & & \cellcolor{lightblue} + & \\ 
            \hline
            Number of unique words & \cellcolor{lightblue} + & & \cellcolor{lightblue} + & \\ 
            \hline
            Number of prompts & & & & \\ 
            \hline
            Average prompt rating & & & & \\ 
            \hline
            Average between-prompt similarity & \cellcolor{lightblue} + & & \cellcolor{lightblue} + & \\ 
            \hline
            Response reading time & \cellcolor{lightblue} + & & \cellcolor{lightblue} + & \\ 
            \hline
            Response length/reading time & \cellcolor{lightblue} + & & \cellcolor{lightblue} + & \\ 
            \hline
        \end{tabular}
    \end{subtable}

    \vspace{3mm} 

    \begin{subtable}{\textwidth}
        \centering
        \caption*{Behavioral Differences Across Frequencies of Virtual Assistant Usage Under Varying Task Types}
        \begin{tabular}{p{4cm}|c|c|c|c}
            \hline
            \textbf{} & \textbf{Choice-fixed} & \textbf{Prioritize-fixed} & \textbf{Choice-open} & \textbf{Prioritize-open} \\ 
            \hline
            Prompt formulation time & & & \cellcolor{lightblue} + & \\ 
            \hline
            Prompt length & & & \cellcolor{lightblue} + & \\ 
            \hline
            Number of unique words & \cellcolor{lightblue} + & & \cellcolor{lightblue} + & \\ 
            \hline
            Number of prompts & & & & \\ 
            \hline
            Average prompt rating & & & \cellcolor{lightblue} + & \\ 
            \hline
            Average between-prompt similarity & & & \cellcolor{lightblue} + & \\ 
            \hline
            Response reading time & & & & \\ 
            \hline
            Response length/reading time & & & \cellcolor{lightblue} + & \cellcolor{lightblue} + \\ 
            \hline
        \end{tabular}
    \end{subtable}
\label{tab:behavior_experience_task}
\end{table}
\endgroup

Going beyond prompting across different task types, the results in Table \ref{tab:behavior_experience_task} illustrate how the relationship between prior system usage and user behavior varies across different task types, revealing the interaction between functional fixedness and task constraints. Notably, the results show that frequent ChatGPT users showed a negative correlation with prompt similarity in choice-open and prioritization tasks, suggesting that they tended to modify their prompts more dynamically in these structured decision-making contexts, compared to users with less chatbot experiences. This implies that experienced ChatGPT users are more flexible in adapting their strategies when faced with predefined choices and ranking tasks, likely because they have internalized the system’s affordances and understand how to navigate multi-turn interactions. Also, users with richer prior experiences tend to spend less time on prompt formulation and system response reading. 

With respect to search engine and virtual assistant experiences, the effects of functional fixedness were more pronounced and task-dependent. Frequent searchers demonstrated significantly higher prompt formulation times, longer prompt lengths, and a greater number of unique words in choice-making tasks, indicating that they took a more deliberative and structured approach when required to compare and rank options. This suggests that they carried over their traditional search tactics, treating these tasks as requiring comprehensive information retrieval rather than dynamic conversation. Similarly, frequent virtual assistant users exhibited higher prompt formulation times and longer prompts in choice-open and prioritize-open tasks, reflecting their tendency to structure interactions in a rigid, command-driven manner, even in contexts that called for open-ended exploration. These findings address RQ2 by showing that functional fixedness is not only shaped by prior system use but may also be moderated by task factors. Frequent search engine and virtual assistant users, in particular, appear to struggle with more flexible task structures, indicating that system interventions, such as adaptive task framing or real-time scaffolding, could help guide these users toward more effective engagement with LLM-based chat systems.

\subsection{RQ3: Impact of In-Situ Expectation Confirmation States on Chat Interactions}

\subsubsection{Impact of Expectation Confirmation on Chat Interactions}
\begin{figure}[h]
    \centering
    \includegraphics[width=0.8\textwidth]{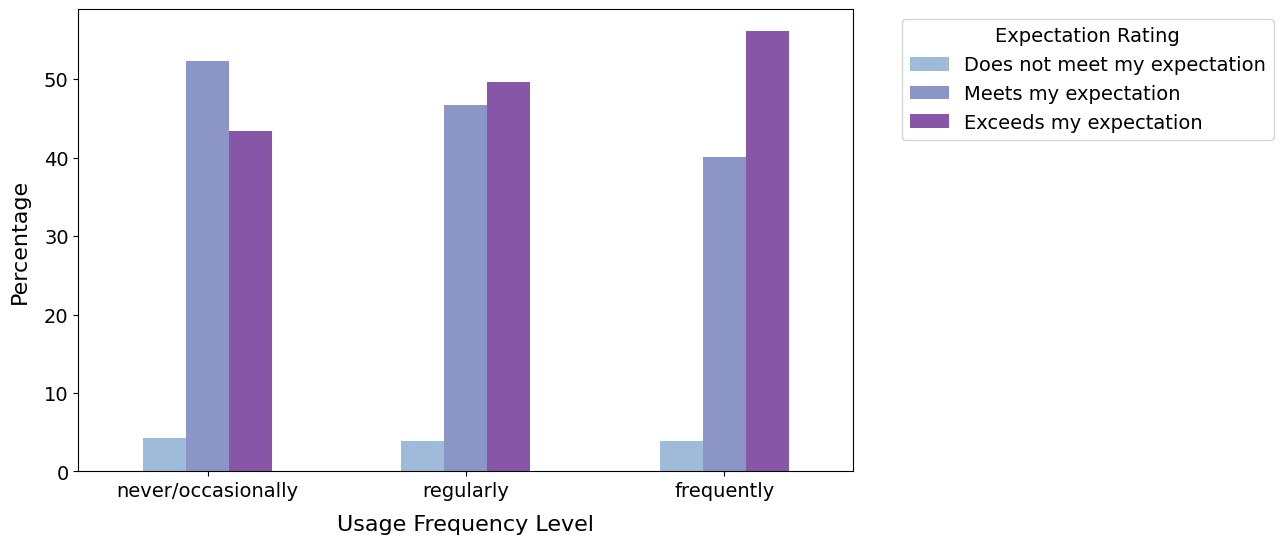}
    
    \setlength{\belowcaptionskip}{0pt} 
    \caption{Percentage of Different Rating Levels by ChatGPT Usage Frequency}
    \label{fig:chatgpt_rating}
    \vspace{3mm}
    \end{figure}
The results in Figure 5 and Table 9 illustrate how users' expectation confirmation states influence their subsequent prompts, shedding light on how functional fixedness may be disrupted or reinforced during interactions. Figure 5 indicates that users with frequent ChatGPT experience were more likely to rate system responses as meeting or exceeding their expectations compared to non-frequent users, indicating that familiarity with LLMs leads to better alignment between user expectations and system performance. In contrast, users with little to no prior ChatGPT experience more frequently rated responses as failing to meet expectations, suggesting a greater mismatch between their preconceived notions and the system’s actual capabilities. This discrepancy highlights the role of prior exposure in shaping expectation alignment and suggests that non-frequent users may be more prone to functional fixedness, struggling to adapt when their assumptions about the system prove inaccurate.  

\begin{table}[t]
    \centering
    \caption{Impact of Current Expectation Confirmation State on the Next Prompt}
    \label{tab:expectation-impact}
    \footnotesize 

    \begin{tabular}{@{}p{0.4\textwidth}cc@{}} 
        \toprule
        \textbf{Prompt Behavior} & \textbf{Kruskal-Wallis test} & \textbf{Pairwise test with corrections} \\ 
        \midrule
        \textbf{Prompt length} & \textbf{10.15**} & \textbf{NM > M**, NM > E**} \\ 
        \textbf{Number of unique words} & \textbf{13.25**} & \textbf{NM > M**, NM > E**} \\ 
        Prompt formulation time & 0.12 &  \\ 
        Response reading time & 5.29 &  \\ 
        Response length/reading time & 5.21 &  \\ 
        \bottomrule
    \end{tabular}

    \vspace{2mm} 
    \footnotesize
    \textit{Note: *: p<0.05. **: p<0.01. \textbf{NM}: does not meet expectation; \textbf{M}: meet expectation; \textbf{E}: exceeds expectation.}
\end{table}

Table 9 provides further insights into how unmet expectations influence user adaptation. When a system response did not meet expectations, users tended to issue longer prompts (p < 0.01) with more unique words (p < 0.01) in their subsequent queries, indicating an effort to refine or clarify their requests. However, there was no significant effect on prompt formulation time, suggesting that users adjusted their queries quickly rather than hesitating or overthinking. This pattern suggests that while unmet expectations can drive users to modify their approach, the extent of their adaptation may still be constrained by functional fixedness, as users may rely on elaboration rather than fundamental changes in interaction strategy. These findings speak to RQ3 by demonstrating that expectation mismatches serve as a trigger for behavioral adaptation, but users may need additional system support, such as proactive reformulation suggestions or contextual feedback, to move beyond rigid prompting and response evaluation skills and fully explore LLM capabilities.

\subsubsection{Impact of Expectation Confirmation on Whole-Conversation Experience}
With respect to whole-conversation evaluation, Table 10 presents correlations between users’ expectation confirmation states and two key outcome measures: whole-session user satisfaction and self-reported task success. The results indicate that the percentage of prompts exceeding expectations is positively correlated with both session satisfaction ($\rho$ = 0.169, p < 0.01) and perceived task success ($\rho$ = 0.15, p < 0.01). Similarly, the percentage of prompts meeting or exceeding expectations shows a weaker but still significant correlation with session satisfaction ($\rho$ = 0.142, p < 0.01), while the percentage of prompts that do not meet expectations exhibits a negative correlation with satisfaction ($\rho$ = -0.143, p < 0.01). It worth noting that the expectation states of the first and last prompts in a session are both positively associated with user satisfaction, suggesting that early experiences set the tone for user perceptions, and final interactions may reshape overall impressions. This finding is also aligned with the phenomena of anchoring bias and recency effect in session evaluations~\cite{liu2020investigating, zhang2020cascade, liu2018towards}. Additionally, average prompt formulation time has significant correlation with user satisfaction ($\rho$ = 0.192, p < 0.01) and task success ($\rho$ = 0.169, p < 0.01), indicating that users who take more time crafting their queries tend to report higher satisfaction levels.

These findings help answer RQ3 by showing that expectation confirmation significantly influences user satisfaction and perceived task success. When users experience more exceed-expectation moments, they are more likely to have a positive evaluation of the session and experience a higher level of satisfaction~\cite{dan2016measuring, wang2023investigating, liu2019investigating}, reinforcing the importance of understanding and aligning user expectations in LLM-enabled interactions. The positive association between prompt formulation time and satisfaction further suggests that users who invest cognitive effort in refining their queries may develop a stronger sense of control and engagement, leading to better experience. These results highlight the need for system designs that support expectation calibration, such as adaptive feedback mechanisms or interface cues that help users set more realistic expectations and encourage iterative query refinement, ultimately mitigating the negative effects of functional fixedness.

\begin{table}[htbp]
    \centering
    \caption{}
    \label{tab:expectation-correlation}
    \footnotesize

    \begin{subtable}{\textwidth}
        \centering
        \caption*{Correlation Between Expectation Confirmation States and Whole-Session User Satisfaction}
        \begin{tabular}{@{}p{0.55\textwidth}c@{}}
            \toprule
            \textbf{Expectation and Behavioral Variables} & \textbf{Spearman’s rank Correlation $\rho$} \\ 
            \midrule
            Percentage of exceed-expectation prompts & \textbf{0.169**} \\ 
            Percentage of meet- and exceed-expectation prompts & \textbf{0.142**} \\ 
            Percentage of not-meet-expectation prompts & \textbf{-0.143**} \\ 
            Expectation state of first prompt in the session & \textbf{0.159**} \\ 
            Expectation state of last prompt in the session & \textbf{0.147**} \\ 
            Average prompt formulation time & \textbf{0.192**} \\ 
            Total number of prompts & 0.002 \\ 
            \bottomrule
        \end{tabular}
    \end{subtable}
    \begin{flushleft}
    \hspace{1cm}
    \textit{Note: *: $p<0.05$. **: $p<0.01$.}
    \end{flushleft}

    \begin{subtable}{\textwidth}
        \centering
        \caption*{Correlation Between Expectation Confirmation States and Self-Reported Level of Task Success}
        \begin{tabular}{@{}p{0.55\textwidth}c@{}}
            \toprule
            \textbf{Expectation and Behavioral Variables} & \textbf{Spearman’s rank Correlation $\rho$} \\ 
            \midrule
            Percentage of exceed-expectation prompts & \textbf{0.15**} \\ 
            Percentage of meet- and exceed-expectation prompts & 0.04 \\ 
            Percentage of not-meet-expectation prompts & -0.045 \\ 
            Expectation state of first prompt in the session & 0.081 \\ 
            Expectation state of last prompt in the session & 0.078 \\ 
            Average prompt formulation time & \textbf{0.169**} \\ 
            Total number of prompts & -0.001 \\ 
            \bottomrule
        \end{tabular}
    \end{subtable}

\end{table}

\section{Discussion}
Our findings provide a comprehensive response to the proposed RQs by empirically demonstrating how functional fixedness manifests at multiple levels in LLM-enabled chat interactions. Regarding \textbf{RQ1}, we established that users’ pre-chat intent-based expectations are significantly influenced by their prior experiences with different digital tools, including search engines, virtual assistants, and ChatGPT-like systems. These prior interactions act as cognitive anchors that shape users’ assumptions about what an LLM can accomplish, often constraining their willingness to experiment beyond familiar patterns. In addressing \textbf{RQ2}, our results reveal that these pre-existing expectations directly influence users’ prompting behaviors. Users with extensive search engine experience tend to frame their prompts in structured, keyword-driven formats, while frequent ChatGPT users exhibit greater fluency in formulating iterative, adaptive queries. Virtual assistant users, by contrast, rely on directive, command-like interactions, which reinforces a more rigid and transactional approach to conversational AI. However, our investigation into \textbf{RQ3} uncovers an important dynamic: while functional fixedness initially limits users’ ability to engage in more flexible interactions, unmet expectations can trigger healthy "frictions" in human-AI interaction~\cite{shah2024envisioning, chen2024exploring, naiseh2021nudging} and motivates chat strategy adaptation~\cite{galland2022adapting}. Some users respond to expectation violations by increasing linguistic diversity in their prompts, while others struggle to break away from pre-established interaction strategies, which suggests that interventions designed to facilitate strategic adaptation could play a crucial role in mitigating functional fixedness. These findings collectively highlight the nuanced relationship between prior experiences, expectation formation, and behavioral adaptation, providing an empirically grounded framework for understanding how users approach generative information search. Note that this study investigates users' functional fixedness in their interactions with ChatGPT, and examines how their chat strategies are connected to and fixated by their prior experiences with different types of interactive information systems. Future research can build on our study design and test our intent typology and findings regarding users' functional fixedness and expectations on other LLMs, generative AI systems, as well as motivating tasks. 

This study makes a unique contribution by demonstrating how functional fixedness affects not only user expectations but also real-time interaction patterns. While previous work on cognitive biases in IR has primarily examined how users evaluate retrieved results~\cite[e.g.,][]{white2013beliefs, wang2024cognitively, draws2021not}, our research extends this discussion to the act of prompt formulation. We show that functional fixedness is embedded in the way users structure their prompts, from the length and linguistic composition of their queries to the degree of variation they exhibit over successive turns. More importantly, our findings reveal that users’ ability to break free from entrenched patterns is not uniform but contingent on the type of prior exposure they have had. Frequent ChatGPT users, while generally more adaptive, still exhibit implicit constraints in their interactions, such as reduced reliance on contextual cues and a preference for concise, direct prompts. Conversely, users with rich search experiences struggle with dynamic adaptation and tend to make incremental modifications to their queries rather than exploring alternative ways of engaging with the system. By uncovering these behavioral tendencies, our study provides crucial insights for the design of LLM-enabled chat systems, emphasizing the need for \textit{interactive scaffolding} that encourages users to adopt a more exploratory and iterative approach to AI-assisted conversational search. Rather than treating functional fixedness as an immutable constraint, our results suggest that it can be actively shaped by unmet expectations and system affordances, which opens up new possibilities for customized and intervention-driven user education that fosters greater flexibility. For future implementations, interactive scaffolding can play a pivotal role in mitigating biases and functional fixedness by dynamically guiding users toward more flexible and exploratory prompt strategies. This could be achieved through adaptive prompt suggestions, real-time feedback loops that highlight alternative phrasings, or system-initiated prompts that encourage users to reconsider implicit constraints in their queries~\cite{krasakis2023contextualizing, sharma2024hybrid, brade2023promptify}. Additionally, multi-modal affordances, such as visual query expansion or interactive example-based demonstrations, can help users recognize diverse ways of structuring their prompts and refine their search strategies iteratively. By embedding these scaffolding mechanisms into conversational interfaces, LLMs can not only enhance users’ interaction performance but may also serve as cognitive training tools that foster long-term adaptability in human-AI interaction.

This study also contributes to the methodological landscape by developing a structured approach for measuring functional fixedness in LLM interactions at scale. Existing research on cognitive biases in online information seeking has largely relied on observational studies or retrospective surveys~\cite{azzopardi2021cognitive, liu2023behavioral}, limiting the ability to capture how biases evolve in real-time user interactions. In contrast, our study employs a controlled experimental design that systematically captures pre-chat expectations, in-situ adaptation strategies, and post-task assessments, allowing us to track how users’ cognitive constraints manifest throughout the interaction process. By integrating expectation confirmation states as a key moderating factor, we demonstrate that functional fixedness is not a static condition but one that fluctuates based on users’ interactions with the system. This methodological innovation provides a replicable framework for future research seeking to examine how cognitive biases shape LLM-based interactions. Moreover, our findings highlight the value of behavioral markers, such as prompt similarity scores and linguistic shifts, as indicators of functional fixedness, offering a new lens through which researchers and engineers can address implicit cognitive barriers in conversational AI. The empirical findings above not only deepen our theoretical understanding of functional fixedness but also inform practical interventions that can be embedded into LLM interfaces to promote more effective user engagement.  

Despite the strengths of this study, there are limitations that warrant further exploration in future research. One limitation is that our participant pool was drawn from a crowdsourced population, which, while diverse, may not fully capture the behaviors of expert users or professionals engaging with LLMs in high-stakes decision-making contexts. Besides, our crowdsourcing study only attracted a fairly small percentage of older adults. Since older adults are more likely to fall behind in technology adoption and be negatively impacted by digital divide~\cite{zhang2025falling, chu2022digital}, future research should identify and mitigate the effect of functional fixedness and cognitive biases in the LLM interactions of older adults and other understudied or vulnerable populations. In addition, it is not appropriate to assume that functional fixedness always negatively affect interaction processes or task outcome. For instance, under some routine tasks, it is possible that fixed approaches may improve productivity in task completion by streamlining problem-solving activities and improving consistency between tasks. Also, fixated approaches can reduce the cognitive load by limiting the number of potential solutions one needs to consider, which makes decisions faster and less mentally taxing. Thus, it is critical to develop and evaluate \textit{functional fixedness assessment} models and adaptively recommend suitable interventions and in-situ user education when the model detects a potentially negative impact of functional fixedness (e.g., limiting exploration and innovation in creative and complex tasks, restricting users to static and routine usages of a flexible new system). 

Future studies should examine functional fixedness in specialized domains such as medical, legal, or technical search, where users may exhibit different forms of cognitive constraints shaped by professional training and domain expertise~\cite{goldstone2010domain}. Researchers should also move beyond ChatGPT to explore a broader spectrum of LLMs and \textit{multi-modal} AI systems, and assess their differential impacts on user cognition, decision-making, and interaction patterns. Given the rapid evolution of generative AI, future research can investigate how model architecture, training data composition, and fine-tuning strategies influence response quality, bias propagation, and trust calibration~\cite{chen2024more, steyvers2024calibration}. Additionally, comparative studies across multiple AI-driven conversational agents can provide deeper insights into system-specific affordances and limitations, shedding light on how users adapt their prompting behaviors and reliance on these tools over time. Finally, interdisciplinary approaches integrating cognitive science, human-computer interaction, and ethics can help design frameworks that ensure AI systems align more effectively with user needs while mitigating risks associated with misinformation, over-reliance, and automation bias. Additionally, our study focuses on short-term interactions and offers a snapshot of how users adapt within a single session. A longitudinal approach would allow for a more nuanced understanding of how functional fixedness evolves over repeated exposures to LLMs, revealing whether users naturally overcome cognitive constraints over time or whether persistent intervention is necessary to address the potential bottleneck in creative interactions. Another avenue for future work is the design and evaluation of targeted system interventions that can actively mitigate biases and functional fixedness~\cite{mccaffrey2012innovation, weatherford2021using}. While our findings suggest that expectation violations can sometimes serve as a catalyst for adaptation, it remains unclear which specific types of system nudges, such as affordance signaling, real-time feedback, or guided prompting suggestions, are most effective in encouraging more flexible user behaviors. Future research should experiment with different intervention strategies and also explore LLM's \textit{Theory of Mind} (ToM)-like capabilities to determine how LLM-enabled systems can best support users in overcoming cognitive biases and constraints and engaging in richer interactions~\cite{kim2025hypothesis, strachan2024testing, kosinski2024evaluating}. Built on our findings, subsequent studies can develop more refined approaches for enhancing the usability and effectiveness of LLM-based conversational search.

\section{Conclusion}
This study advances the understanding of how functional fixedness influences users’ interactions with LLM-enabled chat search, offering novel insights into the interplay between users’ pre-chat expectations, their prior experiences with different digital tools, and their ability to adapt prompting strategies, especially in response to in-situ expectation violations. Differing from previous works that have primarily examined cognitive biases in static search interactions~\cite[e.g.,][]{eickhoff2018cognitive, scholer2013effect}, our work systematically investigates functional fixedness in dynamic, multi-turn chat search environments. Through a crowdsourcing study with 450 participants, we provide empirical evidence that users’ prior experiences with search engines, virtual assistants, and chatbots create distinct cognitive constraints that shape their prompting behaviors and affect their ability to explore the full capabilities of LLMs in interactive chat sessions. Our findings highlight that search engine users exhibit rigid, structured prompting patterns, while virtual assistant users maintain directive, command-like interactions, both limiting their adaptability in chat-based interactions. Frequent ChatGPT users, while more adaptive, still exhibit entrenched habits that constrain linguistic flexibility. By demonstrating that expectation violations can serve as a trigger for behavioral shifts, this study identifies key leverage points for mitigating functional fixedness through system design. These insights have implications for the development of user-centered LLM applications, emphasizing the need for \textit{expectation-aware} interfaces that encourage exploratory engagement. Beyond scientific research, our findings are particularly relevant for industry applications in search, conversational AI, and digital assistants, where optimizing user adaptability is critical for maximizing the utility of interactive systems~\cite{payne2022adaptive}. Designing intelligent chat systems that accurately differentiate positive and negative cases of functional fixedness and actively guide users toward flexible, context-aware prompting strategies could significantly enhance the usability of LLM-enabled chat applications in domains such as customer support, knowledge discovery, and creative problem-solving.  

While our study provides a foundational understanding of functional fixedness in LLM interactions, future research should expand on our findings by exploring interventions that actively support users in overcoming cognitive constraints. One promising avenue is to develop chat interfaces that incorporate real-time feedback mechanisms to help users recognize and adjust fixed interaction patterns. Adaptive prompting recommendations, interactive tutorials, and scaffolding techniques could further assist users in refining their engagement strategies over time. Additionally, longitudinal studies are needed to examine whether functional fixedness naturally diminishes with repeated exposure to LLM interactions or whether persistent intervention is required to facilitate long-term adaptability. Beyond general users, future research should also investigate how functional fixedness manifests in expert and professional settings, where domain-specific knowledge and the complexity of specialized tasks may introduce additional layers of cognitive rigidity. For example, examining how legal professionals, medical practitioners, or educators interact with LLM-enabled search systems could provide insights into whether expertise amplifies or mitigates functional fixedness. Finally, system-level studies should explore how LLMs themselves contribute to reinforcing or alleviating functional fixedness through their response structures, affordance signaling, and interaction histories. By addressing these open questions, future research can build on our findings to develop LLM applications that not only deliver high-quality responses but also facilitate more cognitively flexible, productive, and inspiring human-AI interactions.

\section{Acknowledgment}
This work is supported by a faculty research award from Microsoft and an internal research grant from the University of Oklahoma Office of the Vice President for Research Partnerships.
\bibliographystyle{ACM-Reference-Format}
\bibliography{main}

\appendix
\newpage
\section*{Appendix}
\begin{table}[h]
    \centering
    \footnotesize
        \caption{Predefined Linguistic Marker Categories and Representative Terms Used in Analysis}
    \renewcommand{\arraystretch}{1.2}
    \begin{tabular}{lp{10cm}}
        \toprule
        \textbf{Marker Category} & \textbf{Representative Terms} \\
        \midrule
        \textbf{Politeness Markers} & please, thank, thank you, excuse me, pardon, pardon me, sir, I am sorry, could you, would you \\
        \textbf{Hedge Words} & maybe, sort of, not sure, kind of, think, I believe, I guess, may, suggest, likely, seems, possibly, tend to, could be, might, relatively, generally \\
        \textbf{Conversational Fillers} & um, uh, you know, like, well, so, I mean, actually, literally, basically, alright, hmm, okay, anyways, right, I suppose, kind of, sort of, you see \\
        \textbf{Discourse Markers} & however, but, in fact, I mean, nevertheless, nonetheless, on the other hand, though, even so, regardless, whereas, still, alternatively \\
        \textbf{Deictic Words} & this, that, these, those, here, there, now, then, such, above, below, over there, over here, next, previous, earlier, later, before, after \\
        \bottomrule
    \end{tabular}
    \label{tab:linguistic_markers}
\end{table}

\begin{tcolorbox}[
    colback=gray!10,     
    colframe=gray!80,    
    title=Prompt Classification Template,
    fonttitle=\bfseries,
    breakable
]
\footnotesize 
You are a specialized classifier designed to analyze and categorize prompt interactions. In this study, participants were given a task and engaged with ChatGPT using various prompting techniques. Your objective is to examine the provided data and classify each participant’s prompt into one of the following categories based on its characteristics:

\textbf{1. Logical and Sequential Processing}  
– Techniques that break down complex reasoning tasks into structured, step-by-step components.  
– \textit{Examples:} Chain-of-Thought, Skeleton-of-Thought prompting.

\textbf{2. Contextual Understanding and Memory}  
– Techniques that rely on recalling past interactions or context to generate coherent and relevant responses.  
– \textit{Examples:} Conversational prompting, Socratic prompting.

\textbf{3. Specificity and Targeting}  
– Techniques that focus on precisely guiding the model by specifying details, formats, or objectives.  
– \textit{Examples:} “Show-me vs. Tell-me” prompting, Target-your-response prompting.

\textbf{4. Meta-Cognition and Self-Reflection}  
– Techniques that encourage the model to introspect, evaluate its responses, or refine its output through self-assessment.  
– \textit{Examples:} Self-reflection prompting, Meta-prompting.

\textbf{5. Directional and Feedback}  
– Techniques that provide explicit instructions or feedback to steer the model’s output, either by guiding its thought process or refining previous responses.  
– \textit{Examples:} Responsive feedback prompting, Directional stimulus prompting.

\textbf{6. Creative and Generative}  
– Techniques designed to inspire creativity, exploration, or content generation in tasks such as storytelling, creative writing, or controlled language generation.  
– \textit{Examples:} Ambiguous prompting, Grammar correction, Constrained vocabulary prompting.

\textbf{7. Unrelated}  
– Use this category only if, after thorough analysis, the prompt does not clearly fit any of the above categories. Before classifying as \textit{Unrelated}, attempt to reassign it to the closest matching category. If a prompt is truly unclassifiable, provide a brief description of the patterns or characteristics (e.g., incoherent structure, lack of context) that make it unclassifiable.  

\textbf{Classification Instructions:}  
The data you will receive includes:  
• The history of the user’s previous prompt(s).  
• The current prompt that needs to be categorized.  

Note: The current prompt does not need to be related to previous prompts. However, previous prompts may be useful in recognizing patterns, such as identifying a Logical and Sequential Processing style when tasks are consistently broken down into structured steps.

\textbf{Response Format:}  
Your answer should consist of a single category label (1–7) followed by a brief explanation (1–2 sentences) justifying your choice. Ensure that your explanation highlights key aspects of the prompt that align with the chosen category.

\textbf{----------- Classification Data Below -----------}

Previous prompts: \{history of prompts\}

Current prompt: \{current\_prompt\}

Return your classification for Current Prompt as a JSON object with the following format:

\{
    "category": "category",
    "justification": "justification"
\}
\end{tcolorbox}

\begin{table}[h]
    \centering
    \footnotesize 
        \caption{Examples of Categorized Prompts}
    \begin{tabular}{p{4cm}|p{9cm}}
        \hline
        \textbf{Prompt Category} & \textbf{Prompt Examples} \\
        \hline
        Logical and Sequential Processing & Using the information you gave above, how would you rank all three based on energy efficiency, cost, and environmental impact? Then finally, how would you rank them overall. \newline
        Evaluate the feasibility and impact of adopting a carbon-neutral lifestyle, considering changes in diet, transportation, energy use at home, and other relevant aspects. Is it possible for all of us to adopt a carbon-neutral lifestyle? Why? \\
        \hline
        Contextual Understanding and Memory & What about foam? \newline
        Can everybody adopt this lifestyle? \\
        \hline
        Specificity and Targeting & Give me more detail about the real-time monitoring. \newline
        Please provide a three-paragraph, shortened summary of the above discussion to help me review everything in a consolidated manner. \\
        \hline
        Meta-Cognition and Self-Reflection & So let's start with what is your thought about privacy and public safety? \newline
        How did you choose Keto or Vegan diet? \\
        \hline
        Directional and Feedback & Your answer was very large, make it in pointers and keep only important points. \newline
        More details. \\
        \hline
        Creative and Generative & Assume yourself as EVSGPT. You have been asked to evaluate the feasibility of a carbon-neutral lifestyle and its impact. It can involve diet, the transportation we generally use, the energy we use at home, and other factors. \newline
        What is a smart question about space? \\
        \hline
    \end{tabular}
    \label{tab:exampleprompt}
\end{table}

\end{document}